\documentclass[aps,pre,reprint,twocolumn,superscriptaddress,longbibliography,nofootinbib,floatfix,showpacs,numbers,sort&compress]{revtex4-1}

\usepackage{graphicx,color}
\usepackage{amsmath,amssymb}
\usepackage{hyperref}
\usepackage{verbatim}

\usepackage{ucs} 

\hypersetup{pdfpagemode=UseNone}

\long\def\symbolfootnote[#1]#2{\begingroup%
\def\thefootnote{\fnsymbol{footnote}}\footnote[#1]{#2}\endgroup} 
\def\blfootnote{\xdef\@thefnmark{}\@footnotetext}

\definecolor{purp}{rgb}{0.5,0,0.5}
\definecolor{darkgreen}{rgb}{0.2,0.7,0.1}
\definecolor{orange}{rgb}{1,0.5,0.2}
\definecolor{violet}{rgb}{0.7,0,0.7} 

\newcommand{\RefA}[1]{{\textcolor{black}{#1}}}  
\newcommand{\RefB}[1]{{\textcolor{black}{#1}}} 

\newcommand{\be}{\begin{equation}}
\newcommand{\ee}{\end{equation}}
\newcommand{\bi}{\begin{itemize}}
\newcommand{\ei}{\end{itemize}}
\newcommand{\bea}{\begin{eqnarray}}
\newcommand{\eea}{\end{eqnarray}}
\newcommand{\ud}{\mathrm{d}}

\makeatletter
\def\l@subsubsection#1#2{}
\makeatother

\begin{document}

\title{\RefB{Extended PIC schemes for physics in ultra-strong laser fields: review and developments}}

\author{A.~Gonoskov}
\email[]{arkady.gonoskov@chalmers.se}
\affiliation{Department of Applied Physics, Chalmers University of Technology, SE-41296 Gothenburg, Sweden}
\affiliation{Institute of Applied Physics, Russian Academy of Sciences, Nizhny Novgorod 603950, Russia}
\affiliation{University of Nizhni Novgorod, Nizhny Novgorod 603950, Russia}

\author{S.~Bastrakov}
\affiliation{University of Nizhni Novgorod, Nizhny Novgorod 603950, Russia}

\author{E.~Efimenko}
\affiliation{Institute of Applied Physics, Russian Academy of Sciences, Nizhny Novgorod 603950, Russia}
\affiliation{University of Nizhni Novgorod, Nizhny Novgorod 603950, Russia}

\author{A.~Ilderton}
\affiliation{Department of Applied Physics, Chalmers University of Technology, SE-41296 Gothenburg, Sweden}

\author{M.~Marklund}
\affiliation{Department of Applied Physics, Chalmers University of Technology, SE-41296 Gothenburg, Sweden}

\author{I.~Meyerov}
\affiliation{University of Nizhni Novgorod, Nizhny Novgorod 603950, Russia}

\author{A.~Muraviev}
\affiliation{Institute of Applied Physics, Russian Academy of Sciences, Nizhny Novgorod 603950, Russia}
\affiliation{University of Nizhni Novgorod, Nizhny Novgorod 603950, Russia}

\author{A.~Sergeev}
\affiliation{Institute of Applied Physics, Russian Academy of Sciences, Nizhny Novgorod 603950, Russia}
\affiliation{University of Nizhni Novgorod, Nizhny Novgorod 603950, Russia}

\author{I.~Surmin}
\affiliation{University of Nizhni Novgorod, Nizhny Novgorod 603950, Russia}

\author{E.~Wallin}
\affiliation{Department of Physics, Ume{\aa} University, SE--901 87 Ume{\aa}, Sweden}

\begin{abstract}
\RefB{We review common extensions of particle-in-cell (PIC) schemes which account for strong field phenomena in laser-plasma interactions.} \RefA{After describing the physical processes of interest and their numerical implementation, we provide solutions for several associated methodological and algorithmic problems.}  \RefB{We propose a modified event generator that precisely models the entire spectrum of incoherent particle emission without any low-energy cutoff, and which imposes close to the weakest possible demands on the numerical time step. Based on this, we also develop an adaptive event generator that subdivides the time step for locally resolving QED events, allowing for efficient simulation of cascades.} Further, we present a new and unified technical interface for including the processes of interest in different PIC implementations. Two PIC codes which support this interface, PICADOR and ELMIS, are also briefly reviewed. 
\end{abstract}

\maketitle

\tableofcontents
\section{Introduction} 
Particle-In-Cell (PIC) methods have more than proved their usefulness in plasma physics and related fields \cite{dawson.rmp.1983}. The computational efficiency of PIC approaches originates from their description of a plasma as an (optimal) number of super-particles. In terms of particle distributions in six-dimensional phase-space, this approach makes efficient use of features such as i) similar momentum distribution for neighboring points in coordinate space and ii) the minor role of low density, or empty, regions of phase-space.   

While kinetic and Vlasov-equation-based methods usually deal with distribution functions, a conceptual advantage of the PIC approach is that it is based on particle dynamics. It is thus straightforward to relate the PIC approach both to classical mechanics and (within limits which will be described below) to quantum processes described by particle scattering. This allows the PIC approach to account for particle collisions \cite{peano.pre.2009}, ionization \cite{chen.jcp.2013}, radiation reaction \cite{tamburini.njp.2010, chen.ppcf.2011}, quantum effects \cite{nerush.prl.2011, elkina.prstab.2011, sokolov.pop.2011, ridgers.jcp.2014}, and so on.

In this paper we discuss those extensions of standard PIC approaches which are required to study physics in ultra-intense laser fields. Primarily this requires taking into account the quantized nature of emission from charged particles, as well as various mechanisms of electron-positron pair production. Note that despite the low energy of photons in a laser, the approach is not limited to low-energy physics. Not only are particles rapidly accelerated to high energies by intense fields, but the approach accommodates traditional (i.e.\ accelerator based) sources of high-energy particles.

The realisation of large-scale ultra-intense laser facilities~\cite{ELI, XCELS, GekkoEXA} has greatly stimulated interest in simulations of high-field physics, and several modified PIC implementations have already been developed~\cite{nerush.prl.2011, elkina.prstab.2011, sokolov.pop.2011, ridgers.jcp.2014}. \RefB{The processes of high-energy photon emission and electron-positron pair production have been taken into account using probabilistic routines based on QED calculations, as will be reviewed below.} \\

\RefB{A number of methodological and algorithmic problems have however been raised in association with these implementations. For example, existing implementations of stochastic photon emission (the so-called standard event generator based on inverse sampling, and an alternative event generator described in~\cite{elkina.prstab.2011}) can only account for photons having energy above some certain cutoff value. One can of course take this cutoff to lie well below the energy scale of typical interest, namely that required for pair creation.} \RefA{However, the neglected part of the emission spectrum can still give rise to relevant phenomena that significantly impact particle dynamics; one example is classical radiation reaction. Such phenomena can in particular occur during the early stages of laser-particle interactions, in regions where the laser intensity is low, giving a `knock-on' effect for subsequent interactions in the high intensity part of the pulse.}

\RefB{Some other issues are the possibility to neglect certain QED processes, the problem of double-counting radiation, memory overload due to cascades of particle production, control of computational costs for the statistical routines, and rapidly growing demands on the time step due to a necessity in resolving QED events.}\\

In this article, we analyze some of these problems and propose an original set of modifications capable of solving them. We also present a technical unified interface for merging such modifications with arbitrary PIC implementations, and demonstrate how the idea is realized for the codes ELMIS~\cite{gonoskov.phd.2013} and PICADOR~\cite{bastrakov.jcs.2012}. \\

The paper is organised as follows.  In Sections II and III we discuss general methodological aspects. In Sect. IV and V we discuss implementation of emission from charged particles and pair production, respectively. \RefB{In particular, the previously proposed event generators for describing incoherent emission from charged particles are described and analyzed in Section IV.F, whereas our modified event generator and the adaptive event generator are presented and compared in Sections IV.G and IV.H, respectively.} Some methods of controlling the related computational costs are described in Section VI. In Section VII we benchmark our numerical scheme against existing literature results. We conclude in Sect.~\ref{SECTION:CONCLUSIONS}.

In the appendix we describe the PIC-MDK interface used for merging various PIC schemes with the modules which account for QED processes. We also give some details of the modules' implementation and briefly review two PIC codes ELMIS and PICADOR, which support the interface.


\section{Dual treatment of the electromagnetic field}
%
Traditional PIC approaches treat plasmas as ensembles of charged particles moving in electromagnetic (EM) fields defined on a grid~\cite{dawson.rmp.1983}. Current densities also defined on the grid are the sources of the EM-fields. The evolution and interactions of the particles and fields are then described self-consistently by using the classical equations of motion for charged particles and Maxwell's equations for the EM-fields.
 
This approach takes into account only the radiation that can be resolved by the numerical grid, however. The implied restriction on the numerical scheme can be simply formulated in terms of energy: to accurately capture the physics involved, most of the EM-energy must remain within the spectral range resolved by the grid, $\omega < \omega_{grid} = c/\ud x$, for grid size $\ud x$. As the EM-field intensity rises, particles will be accelerated to higher energies, i.e.\ higher gamma factors. This extends the spectral range of synchrotron emission, whose typical frequency scales classically as~\cite[\S74]{LL.V2}
\begin{equation}\label{cl_freq}
	\omega_c = \frac{3eH_{\text{eff}}}{2mc} \gamma^2,
\end{equation}
in which $e$ and $m$ are the electron charge and mass, $\gamma$ is the electron's gamma factor, $c$ the speed of light, and $H_{\text{eff}}$ is the effective magnetic field defined to be that which can cause the same transverse acceleration as experienced by the electron. (The explicit form of $H_\text{eff}$ will be given below.) One might think that this effect requires successively decreasing the grid size of the cells in the PIC code, ruining the possibility for simulating ultra-relativistic physics. Fortunately, the spectral distribution of the EM-energy is fundamentally uneven, occupying two well-separated regions. This motivates the possibility of merging two different approaches to describing the EM-field, as we now explain.

The first region is associated with synchrotron emission from an electron, and follows the classical spectrum~\cite{LL.V2}
\begin{equation}\label{cl_spectrum}
 \frac{\partial I}{\partial \omega} = \frac{\sqrt{3}}{2\pi} \frac{e^3 H_{\text{eff}}}{mc^2} \frac{\omega}{\omega_c}\int^{\infty}_{\omega/\omega_c} K_{5/3}\left(\xi\right) d\xi.
\end{equation}
As is well known, the emission energy is concentrated in the vicinity of $\omega_c$, and decreases like $I \sim \omega^{1/3}$ with decreasing frequency $\omega$. However, in the frequency range of classical plasma physics the emission intensity dramatically increases due to coherency, which enhances the individual incoherent synchrotron emissions by a factor of the number of particles emitting coherently. This number can be roughly estimated as the number of particles within the typical space scale $\lambda^3$ where $\lambda = 2 \pi c/\omega$ is wave length. Thus, if the wave length is larger than the typical distance between particles, we have an additional factor $\sim c^3 N_e \omega^{-3}$, where $N_e$ is the electron density. (Potentially, the particles can emit coherently even when separated by distances larger than the wave length, but this is an exceptional case which requires some external synchronisation.) As a result, for $\omega < \omega_{coh} = c N_e^{1/3}$ the emission intensity increases with decreasing $\omega$ as $I \sim \omega^{-8/3}$ \footnote{\RefB{The power-law decay for the spectrum coincides with that obtained from the ROM model for HHG via irradiation of a semi-infinite plasma~\cite{baeva.pre.2006}, though we obtain it here from completely different arguments that are not related to any specific geometry.}}. This rise continues down to the typical macroscopic scales characterising the system, such as e.g.~the plasma frequency $\omega_p$, laser frequency $\omega_L$, and the frequency $\omega_t = c/R_t$ associated to the typical space scale $R_t$ of the target/process.

For laser-plasma interactions the typical frequency scale lies in the range $10^{14}$ --$10^{17}$~s$^{-1}$, or photon energies of $0.1\,$eV to $100\,$eV. For the typical density of plasma generated from the ionisation of solid targets ($N_e \sim 10^{21}$~cm$^{-3}$) the value of $\omega_{coh}$ can be estimated as $\hbar\omega_{coh} \sim$~keV. To estimate the typical value of $\omega_c$ for the case of significant radiation losses, we can equate the energy emitted by an individual particle during a single laser cycle to the particle's energy of oscillation. For optical laser frequencies this yields $a \sim 100$ as the typical EM-field amplitude where radiation losses become significant. This corresponds to $\omega_c \sim 10^{-21}$~s$^{-1}$ or photon energies of the order of 1~MeV.
\begin{figure}
\centering\includegraphics[width=1.0\columnwidth]{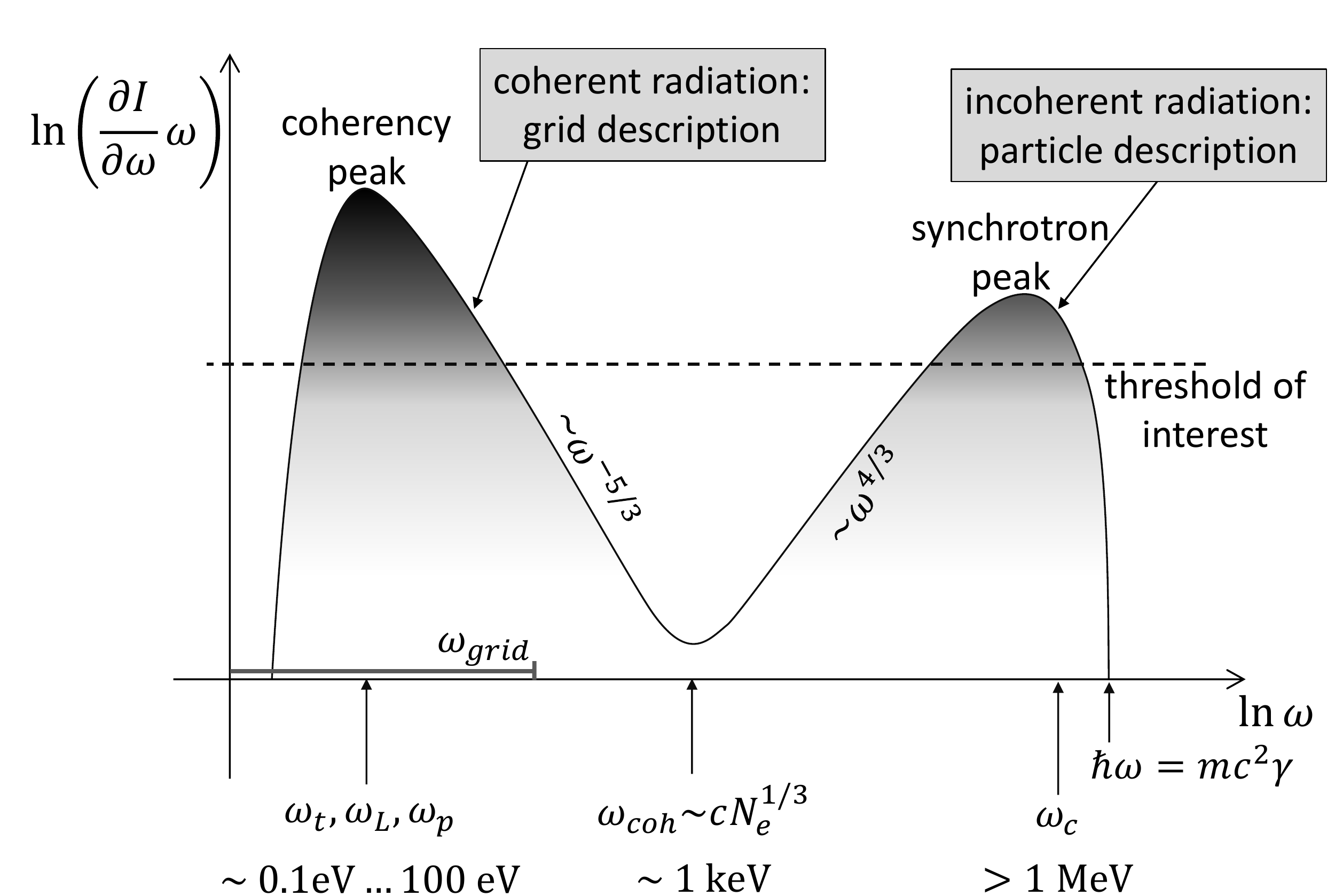}
\caption{Schematic representation of the typical EM-energy deposition.}\label{EM_dep}
\end{figure}

\RefB{Based on the above arguments we illustrate in Fig.~\ref{EM_dep}} the EM-energy distribution using the function $\omega \partial I / \partial \omega$, the integral of which (w.r.t.\ $\log\omega$) gives the emitted energy. Note the log-scales. The figure illustrates the presence and separation of the two regions of EM-energy deposition, which is the basis of the dual treatment of the EM-field described above: the grid approach for low frequency emissions and the particle approach for high-frequency emissions.

\RefB{The incoherent peak is not necessarily separated in case of {\it low} intensities, but then it will contain a negligible part of the emitted energy. Based on the above arguments though, we assert that once the intensity becomes sufficient to convert a notable part of the particle energy into incoherently emitted radiation, then the incoherent peak appears in a well separated spectral region and the dual description of the electromagnetic field becomes possible.}

\RefB{Our arguments so far have been general and have not relied on any specific geometry. To demonstrate the validity of our arguments we now present numerical results for a particular case. We consider the so-called giant attosecond pulse generation in the process of an overdense plasma irradiation by an ultrarelativistic laser in the \textit{relativistic electronic spring} (RES) regime~\cite{gonoskov.pre.2011}. We choose this example because it is a `worst case' scenario; a very wide (far beyond $I \sim \omega^{-8/3}$) spectrum of coherent emission is generated, which means that the two radiation regions could potentially have significant overlap, invalidating our results; we will see that this does not happen.}

\RefB{According to the predictions of RES for optimal conditions, we consider an angle of incidence $\theta = 60^{\circ}$ and relativistic similarity parameter $S = n/a = 0.375$, where $n$ is the plasma density in units of critical density, and $a$ is the radiation amplitude in relativistic units. We have performed PIC simulations for three different intensities with optimal density calculated according to the given $S$. As our aim here is only to verify the predicted form of the radiation spectrum, we turn off pair production, consider the ions to be immobile, assume a sharp drop for the plasma density and consider only one cycle of the incident radiation.}
\begin{figure}
\centering\includegraphics[width=1.0\columnwidth]{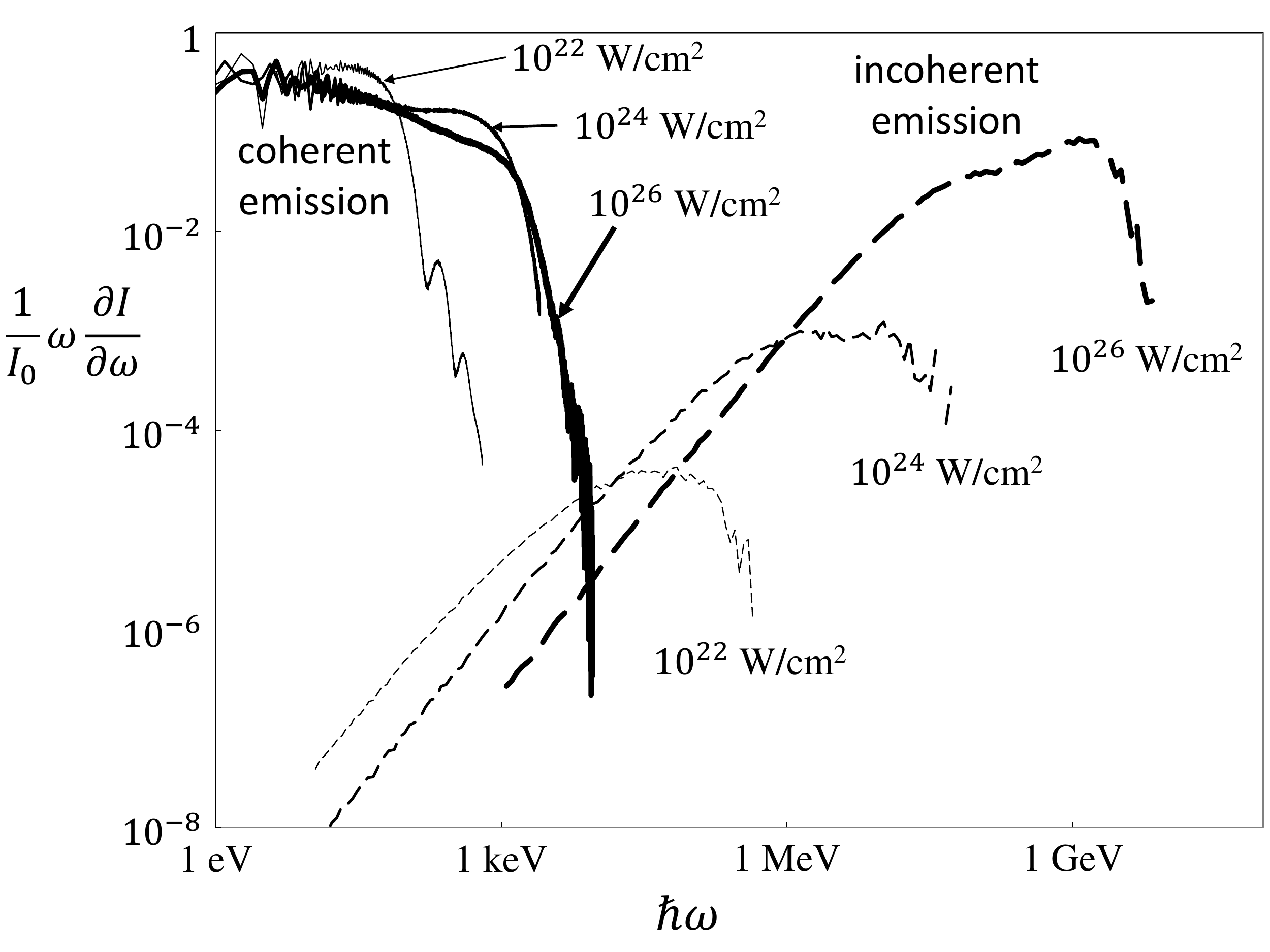}
\caption{\RefB{Spectra of coherent and incoherent emission obtained with PIC simulations for the process of giant attosecond pulse generation in the regime of \textit{relativistic electronic spring} (see \cite{gonoskov.pre.2011} for details).}}\label{spectra}
\end{figure}

\RefB{In Fig.~\ref{spectra} we show the spectra obtained by Fourier transformation of the generated field structure (solid curves) and by accounting for the photons produced by the adaptive event generator described in Sect.~\ref{SECTION:AEG} (dashed curves). The spectral energy density is normalized to the initial energy of the incident radiation $I_0$. As one can see, once the energy of the incoherent emission becomes large enough to require inclusion in the PIC simulation, its spectrum appears in a region well-separated from that of the coherent emission, as predicted. Taking more realistic conditions in our simulation would most probably result in a narrower spectrum of coherent emission, which would match even better with our general conclusions.}


%
\section{Numerical model}
Within the traditional PIC framework, the classical evolution and interplay of fields and particles has been thoroughly studied, see~\cite{dawson.rmp.1983} for reviews. Therefore we focus here on the novel channels and interactions that are opened by laser fields of extreme intensity. We begin with an outline of the numerical model. We then go through which quantum processes are included in the PIC approach, which are neglected, and why, before describing how the included processes are combined into reactions involving many particles. 

\subsection{Implementing quantum interactions in classical PIC}
Classically, the time-evolution of an initial distribution of particles and fields is determined by the appropriate equations of motion. This is straightforwardly implemented in the traditional PIC approach: particle trajectories, for example, are determined and tracked as time evolves in discrete steps.
	
That situation is different in the quantum theory. The concept of a trajectory is neither clear nor necessarily useful (defining a position operator in quantum mechanics is a long-standing issue~\cite{Newton:1949cq}). Some of the most commonly studied objects in quantum field theory, and those which are implemented in PIC codes, are scattering probabilities, that is probabilities for a given asymptotic state (of well separated particles) in the infinite past to evolve into another asymptotic state in the infinite future~\cite{Lehmann:1954rq,Lehmann:1957zz}. Only the initial and final states are specified, not the intermediate dynamics.
	
The incorporation of scattering probabilities into PIC schemes begins with the result, derived and described in~\cite{nikishov.jetp.1964,RitusReview}, that a particle in a high-intensity field sees that field, locally, as a constant, homogenous, plane wave with orthogonal electric and magnetic fields of equal magnitude -- a ``crossed'' field. With this in mind, scattering probabilities are added to PIC schemes as follows.
	\bi
		\item[1.] The probability of a chosen scattering process is calculated assuming the presence of a constant crossed field~$F_c$.
		\item[2.] Such probabilities are infinite, but dividing out the infinite interaction time gives a finite rate $R$.
		\item[3.] $R$ is assumed to give the {\it local} transition rate of the considered process, i.e.\ that occurring at a certain point and time $x^\mu$, and in an arbitrary background field $F$, by replacing the constant field in $R$ with the value of $F$ at the considered spacetime point, i.e.\ $R(F_c)\to R(F(x))$. In short, a locally constant approximation is used. 
		\item[4.] After each time step $\Delta t$ these rates are combined with a statistical event generator in order to decide whether or not a given process occurs, and particles are then added or removed from the simulation as appropriate. Between time steps, particles are propagated forward on their classical trajectories (and fields develop according to their classical equations of motion).
\ei
This model is not a numerical discretization of QED, but rather of classical electrodynamics, into which quantum effects are added by hand. (Contrast with lattice QCD~\cite{PhysRevD.10.2445,Kogut:1979wt}, which is a nonperturbative discretisation of quantum chromodynamics from which one recovers, in the limit of small lattice spacing, continuum QCD.)   As such the model must be tested against known analytic results in order to verify its validity -- for recent tests see~\cite{King:2013zw,elkina.prstab.2011,Harvey:2014qla}. We now describe which scattering processes are included in the numerical model.

\subsection{Included processes}\label{SECT:IN}
The basic interaction vertex in QED connects one (two) incoming to two (one) outgoing particles, real or virtual. One photon and two fermions meet at each vertex. A single vertex cannot describe a scattering process between real particles in vacuum due to momentum conservation. In the presence of a background (e.g.\ an intense laser field) though, it can describe four distinct processes. Two of these are included in the PIC model and two are neglected, as we now describe.

\subsubsection{Photon emission/Nonlinear Compton scattering}\label{NLC1}
\begin{figure}[t!]
\centering\includegraphics[width=1.0\columnwidth]{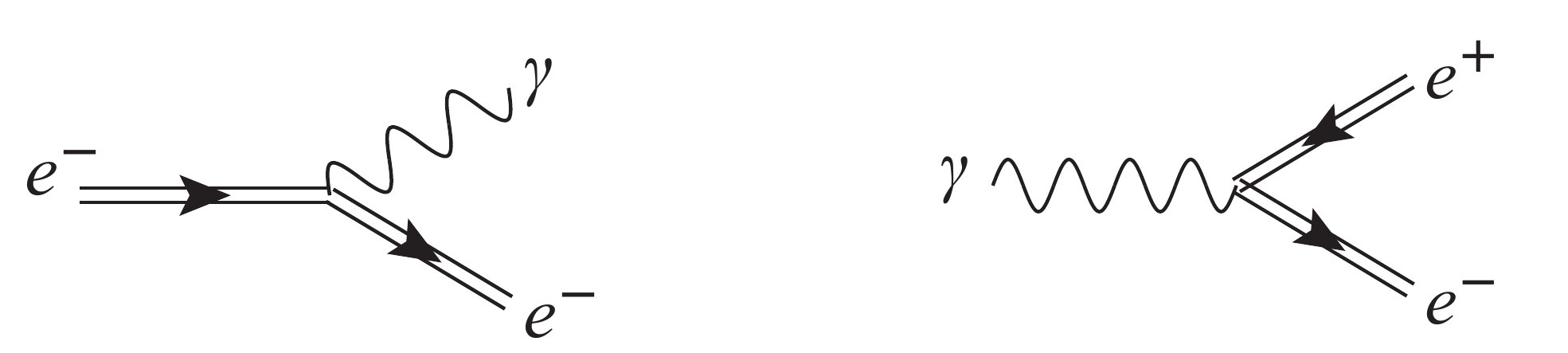}
	\caption{\label{FIG:IN} Feynman diagrams for the two basic processes included in the PIC approach.  Time flows from left to right. A straight line indicates an electron or positron interacting with an external field (e.g.\ an intense laser), and a wavy line is a photon.  {\it Left:} nonlinear Compton scattering. {\it Right:} stimulated pair production.}
\end{figure}

When a charged particle is accelerated by an external field, it emits radiation. This fundamental and familiar process goes under many names, depending on the external field in question; in the case that the electron is excited by a laser and emits a photon it is called `nonlinear Compton scattering'~\cite{nikishov.jetp.1964,nikishov2}. The word `nonlinear' stems from the possibility of the electron absorbing an arbitrary number of laser photons before and after it emits. (The extension to multiple emissions is given below.) Nonlinear Compton scattering is represented by the Furry-Feynman diagram on the left in Fig.~\ref{FIG:IN}.

There is no energy or intensity threshold to overcome in order for this channel to open. The low energy limit of the emitted photon spectrum naturally matches that obtained in (relativistic) classical mechanics, i.e.\ it yields the spectrum of a particle accelerated by the Lorentz force~\cite{LL.V2}. For fast moving electrons with $\gamma\gg 1$ the high-energy part of the emission spectrum is confined to a cone of opening angle $\theta \sim 1/\gamma$, as particles mainly emit forward in the ultra-relativistic limit~\cite{Jackson}. The spectrum in general can exhibit a rich structure strongly dependent on the field configuration. Harmonic structures corresponding to multiple photon absorption can be distinguished by tuning parameters. Nonlinear Compton scattering can be a dominant source of hard photons with an upshifted frequency of $4\gamma^2\times$~the laser frequency; for more details on this and other features of the nonlinear Compton spectrum, see~\cite{nikishov.jetp.1964,nikishov2,Harvey:2009ry,Heinzl:2009nd,Boca:2009zz,Mackenroth:2010jr,Dinu:2013hsd} and references therein.

It is also worthwhile considering the scattered {\it electron} spectrum. Aside from the acceleration caused by the Lorentz force, the momentum of an electron moving through a background field is impacted by recoil from all possible photon emissions, single or multiple~\cite{Higuchi:2005an,DiPiazza:2010mv,DiPiazza:2011tq,Ilderton:2013tb}. Recoil is included unambiguously in QED because momentum is conserved at each interaction vertex: there is always recoil when the electron emits, and this can be seen in the electron spectrum following emission. The classical limit of this recoil describes, of course, the impact of classical radiation reaction on the particle's motion~\cite{Krivitsky:1991vt,Higuchi:2005an,Ilderton:2013dba}.

\subsubsection{Pair production}
High energy photons, such as those generated by nonlinear Compton scattering, can interact with an arbitrary number of laser photons and produce real electron-positron pairs. This generalisation of the Breit-Wheeler process is called `stimulated pair production'. It is a perturbative process in the sense that it is of order $\alpha$ in the interaction between the stimulating photon and the pair (there is a single vertex, see Fig.~\ref{FIG:IN}). However, the interaction between the photon and the background can have both perturbative and non-perturbative dependencies on field strength and kinematics~\cite{Toll:1952rq,Erber:1966vv,Baier:2009it,Heinzl:2006pn}.

Pair production is a threshold process, the probability of which vanishes in the low energy or classical limit. Pair production by two photons in vacuum, for example, has a threshold energy (squared) of $k.k'>2m^2c^4$, where $\{k,k'\}$ are the photon momenta. In an intense laser field, the threshold can be shifted due to both kinematic and intensity effects. For example, for pair creation by $n$ photons of momentum $k$ and one of $k'$ the threshold becomes $nk.k'>2m^2c^4$, leading again to harmonic structure, and in very long pulses intensity effects conspire to raise the pair production threshold from $2m^2c^4$ to $2m^2(1+a_0^2)c^4$, see~\cite{nikishov.jetp.1964,nikishov2}. For pair production in short pulses, see~\cite{Heinzl:2010vg,Nousch:2012xe}.

Stimulated pair production is just one channel by which light can be transformed into matter, others will be discussed below. The production of an electron-positron pair, by any means, requires an energy of at least $2mc^2$, which should be deducted from the energy of seed particles (or, in the case of Sauter-Schwinger pair production, from the EM-field energy). However, in extremely intense fields, the produced electrons and positrons are accelerated rapidly to ultra-relativistic speeds, which requires taking much higher ($\gamma \gg 1$) energy from the EM-field via the channel of ``\textit{classical plasma physics}".  Hence the loss of an energy $2mc^2$ can be neglected; the implied relative error is of order $1/\gamma\ll 1$.

\subsection{Negligible processes}
There are two basic processes, or energy conversion channels, that are weak enough to be neglected in the numerical model: these are the channels of pair annihilation to one photon, and absorption of one photon by an electron or positron, see Fig.~\ref{FIG:OUT}. We explain here the reasons for these simplifications.

\subsubsection{Pair annihilation to one photon}
 \begin{figure}[t!]
\centering\includegraphics[width=1.0\columnwidth]{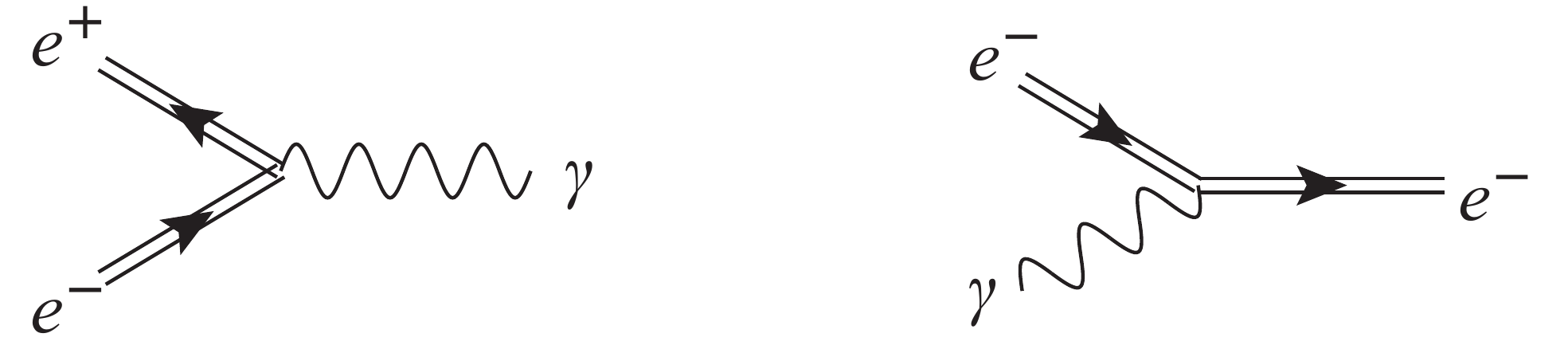}
	\caption{\label{FIG:OUT} Feynman diagrams for the two basic processes neglected in the PIC approach: pair annihilation to one photon  (left) and single photon absorption (right). Time flows from left to right.}
\end{figure}
The annihilation of an electron-positron pair into two photons is an elementary and well-understood process in QED~\cite{PhysRevD.34.3286}. The presence of the laser background opens a new channel, namely annihilation to a single photon, see Fig.~\ref{FIG:OUT}.  Although related by crossing symmetry to the processes above, one-photon annihilation can be neglected, for the following reasons.
 

In constant crossed fields, the annihilation rate is suppressed by an infinite volume factor compared to the processes above~\cite{nikishov.jetp.1964,nikishov2}. This is due to energy-momentum conservation allowing at most a single possible four-momentum for the produced photon. In other words, and unlike in the cases of nonlinear Compton scattering and pair production, the phase space of the final state is determined entirely by the kinematics of the incoming particles, and collapses to a single point.  This remains true even when the calculation is extended to include finite pulse duration and time-dependent structure~\cite{Voroshilo}.

In order for the one-annihilation channel to open, then, the kinematics of the incoming particles must be fine-tuned so that precisely the correct point of phase space is accessible. In the high intensity regime, with ultra-relativistic particles, this only happens when the collision angle $\theta$ between the colliding electron-positron pair obeys $\theta\leq 10^{-5}$rad~\cite{Voroshilo}; it is this degree of fine-tuning which makes the process negligible.

\subsubsection{One photon absorption}
While the emission of photons from an electron in a laser field is essential to include, the absorption of (non-laser) photons by the electron can be neglected. The reasoning is the same as for pair annihilation above~\cite{RitusReview}, as can be seen from Fig.~\ref{FIG:OUT}.

\subsection{Higher order processes}\label{SECT:HIGHER}
`Higher order processes' are those which are described by diagrams containing more than one of the basic vertexes shown above, as dictated by the Feynman rules in QED. There are subtleties in the PIC implementation of higher order quantum processes. To illustrate both these and the functionality of the PIC codes it is simplest to focus on a concrete example, and for this we choose the important process of pair production from a seed electron, the diagram for which is shown in Fig.~\ref{FIG:TRIDENT}.  (See~\cite{Hartin,Hu:2010ye,Ilderton:2010wr,Seipt:2012tn,Mackenroth:2012rb,King:2013osa} for recent investigations of various higher order processes.)

\subsubsection{Trident pair production and cascading}
\begin{figure}[t!]
	\includegraphics[width=0.4\columnwidth]{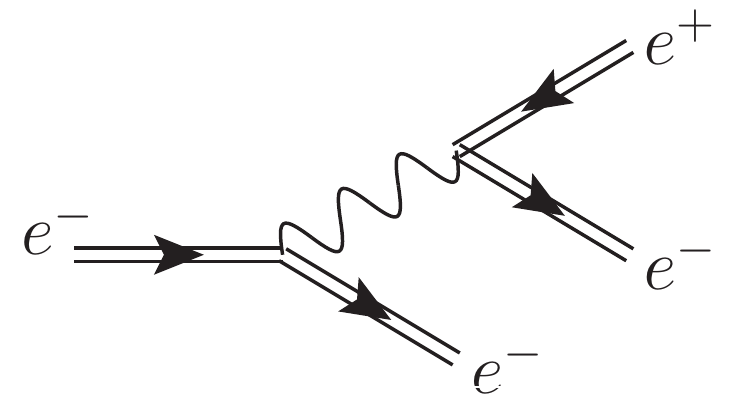}
	\caption{\label{FIG:TRIDENT} Pair production from a seed electron. The intermediate photon may be real or virtual.}
\end{figure}
A seed electron enters an EM-field, radiates a photon, and this photon then produces a pair, see Fig.~\ref{FIG:TRIDENT}. In a PIC simulation, a photon can be emitted via nonlinear Compton at one time step, propagate, and when rates are calculated at the time next time step there is a chance that this photon will produce a pair via stimulated pair production. In this way we build up higher-order process from the basic processes described in Sect.~\ref{SECT:IN}.

At first sight, the PIC description seems to match Fig.~\ref{FIG:TRIDENT} but, in QED, the intermediate photon can be either real or virtual, with both alternatives being captured by and included in Fig.~\ref{FIG:TRIDENT}~\cite{baier72,Ritus:1972nf}. If follows that including the full rate for this process in the same manner as the basic processes above would imply double counting, as the full rate would already allow for the possibility that a real photon is emitted, propagates, and then produces a pair.
It has been found, though, that the contribution from the virtual contribution is usually small in comparison to the case of ``nonlinear Compton $\otimes$ stimulated pair production''. The argument is  then that the full process can be approximated by that part which is mediated by real photons, i.e.\ that part which can be constructed from just the two basic processes in Sect.~\ref{SECT:IN}.

The decomposition of the full process into real and virtual channels has recently been investigated in~\cite{Ilderton:2010wr,King:2013osa} and the process has been analysed in detail for monochromatic fields in~\cite{Hu:2010ye} and for constant fields in~\cite{baier72,King:2013osa}.

\begin{figure}[t!]
	\includegraphics[width=0.7\columnwidth]{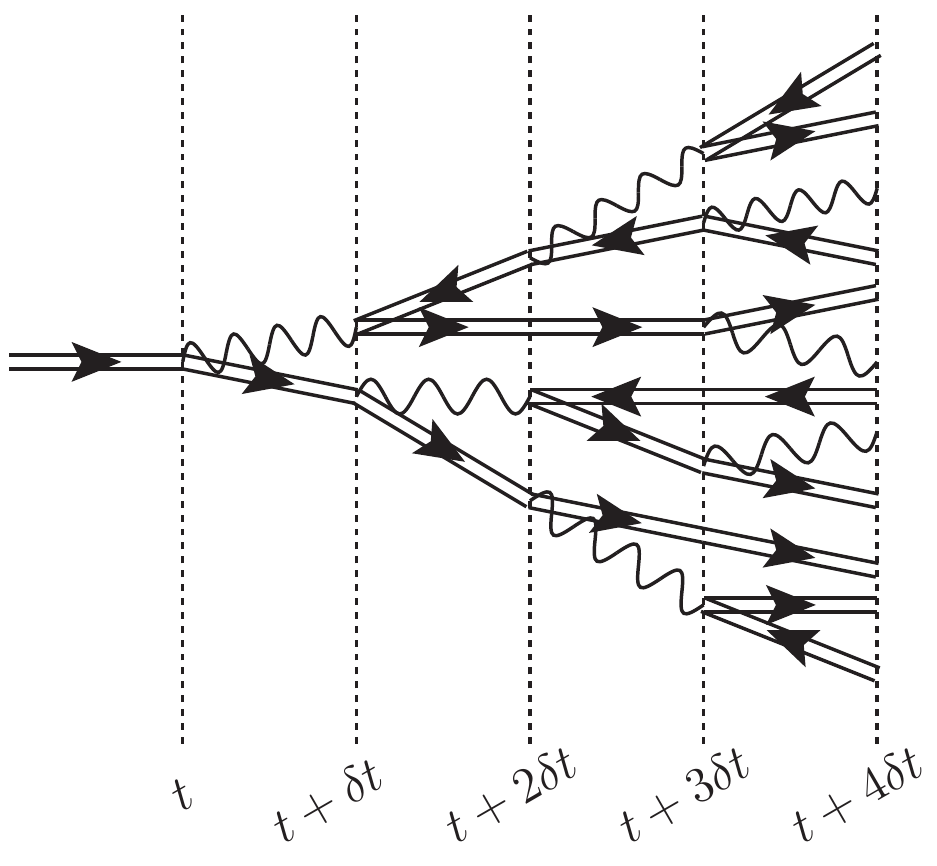}
	\caption{\label{FIG:CASCADE} A (greatly simplified) illustration of cascade formation from a seed electron in a PIC simulation. At each time step, fermions can emit photons (or not) and photons can produce pairs (or not). All particles are real.}
\end{figure}

Repeated nonlinear Compton scattering and pair production events can lead to an avalanche of particle production, or `cascading', see Fig.~\ref{FIG:CASCADE}. The conditions for this are discussed in~\cite{Fedotov:2010ja,elkina.prstab.2011,nerush.prl.2011}, and we will return to cascading in Sect.~\ref{SECT:CASCADE}.

\subsubsection{Multiple photon emission}\label{NLC2}
In the high-intensity regime where $a_0\gg1$ sets the dominant scale, multiple photon emissions from a given fermion are expected to factorize into products of repeated single photon emissions, as for other higher-order processes. The emission of very high energy photons can be included through a synchrotron module, efficiently calculating the spectral properties of the emitted photons~\cite{wallin.pop.2015}.

Two-photon emission has been considered in some detail in the literature and serves to illustrate that the assumptions relevant to the high-intensity regime are not universally applicable. It was found in~\cite{Seipt:2012tn} that for high energy electrons, $\gamma=10^4$, and (by modern standards) low intensities, $a_0 = 1$, that the off-shell channel generated a significant number of photons. This is a situation in which it is not possible to neglect virtual photon contributions. A comparison of the cases $\chi\ll 1$ (negligible quantum effects) and $\chi\gtrsim 1$ (significant quantum effects) in double photon emission is given in~\cite{Mackenroth:2012rb}.

\subsubsection{Neglected higher order processes}

Loop corrections, i.e.\ corrections in higher powers of $\alpha\simeq 1/137$~\cite{Ritus:1972ky}, are neglected in the numerical model. Further, some processes cannot be captured by the model due to the assumptions made; for example, because only photon emission and pair production are included, there is no mechanism by which pairs can annihilate (to any number of photons). It is of course easy to imagine different scenarios in which pair annihilation could be important. Two examples are i) a hohlraum in which we wish to establish long-term equilibrium of the pair plasma, or ii) a very high-density pair plasma where the annihilation rate is of the same scale as the production rate.

\subsection{Summary}
\begin{figure*}[t!]
\centering\includegraphics[width=\textwidth]{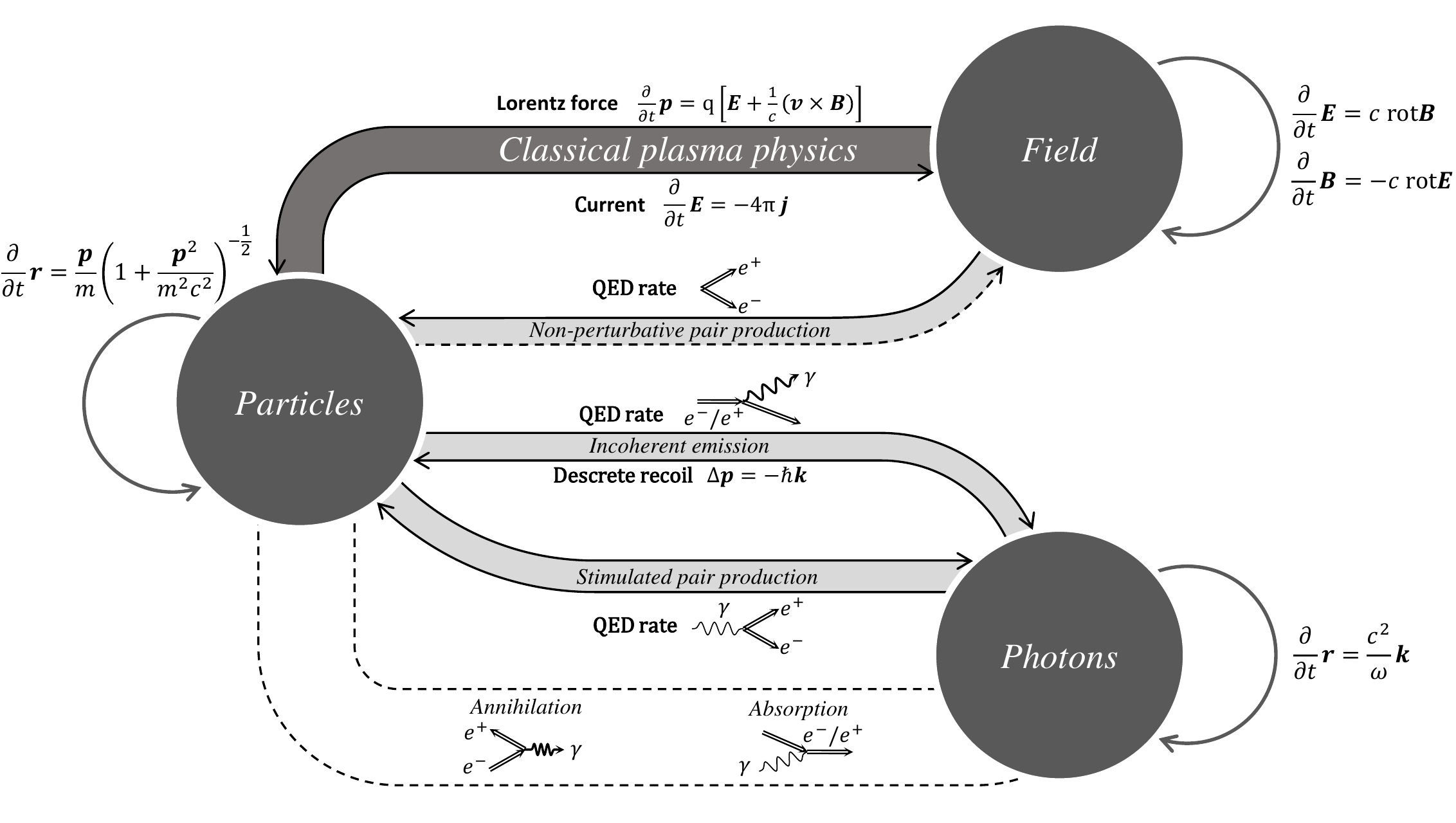}
\caption{\label{scheme} Extension of the PIC approach for taking into account novel channels of energy transformation that could be triggered by laser fields of extreme intensity.}
\end{figure*} 
Despite the common origin for all energy deposition in Fig.~\ref{EM_dep}, we will from here on use the word \textit{field} to mean coherent, low frequency radiation which can be resolved on the simulation grid, and the word \textit{photons} to refer to the incoherent, high frequency radiation given by an ensemble of photons. Using this notation in Fig.~\ref{scheme} we show, schematically, three qualitatively different forms of energy allocation and the possible channels for conversion of energy between them.

The solid lines indicate channels included in the PIC code; these are photon emission and pair production, and higher order processes such as cascades are built up from repeated emission and production events.  The dashed lines indicate, for completeness, the processes which can be neglected. (This is annihilation, absorption and the negligible loss of energy in Sauter-Schwinger pair creation discussed above.) 

We remark that neither spin nor polarisation are included in the code -- for an analysis of photon polarisation effects see~\cite{King:2013zw}.

\section{Incoherent emission}
The emission of classical radiation is well understood as a part of the traditional PIC approach. Here we focus on accounting for incoherent emission of individual photons from electrons and positrons.

\subsection{Problem of double counting}

Accounting for the individual emissions of each photon while simultaneously solving Maxwell's equations implies a double counting, as the current sources of the photons is treated also as a source of coherent emission in the classical equations. However, Fig.~\ref{EM_dep} makes it evident, that the double counting is negligible in terms of energy deposition and thus can hardly affect the macroscopic dynamics of laser-plasma interaction. Indeed, the double counting occurs only in the low-frequency part where the energy contribution of the individual emission is smaller than the one of coherent emission by a factor of number of particles emitting coherently. This number is very large for the typical spectral range of energy deposition for coherent emission.

\subsection{Properties of individual emission}

As estimated above (see Sect.~II), energy loss due to radiation typically impacts particle dynamics significantly for field amplitudes $a \gtrsim 100$. If one is interested in the diagnostics of emission at lower amplitudes one can, assuming the emission does not affect the process, use PIC approach without any modifications and obtain the diagnostic from post processing, calculating the emission integral~\cite{Jackson} over the particles' trajectories (see for example~\cite{chen.prstab.2013}) or using statistical routines~\cite{wallin.pop.2015}. In this article we primary focus on the case of the notable converse effect of emission, thus we assume ultra-relativistic dynamics $\gamma \gg 1$, $a \gg 1$.

To determine emission that an electron produces during one time step of PIC simulation we assume that the electric and magnetic field vary insignificantly during this interval of time. Next, we note, that in ultra-relativistic case emission of a particle is predominantly defined by the transverse acceleration (the longitudinal acceleration has $\gamma^2$ times less contribution to the emission intensity, see \cite{LL.V2}). The emission is determined not by the EM-field itself, but by the transverse acceleration it causes. Thus to determine the emission properties we can use the solution for an electron rotation in a constant uniform magnetic field $H_{\text{eff}}$ whose strength causes a transverse acceleration equal to the one that the electron experiences due to the actual electric and magnetic field. This \textit{efficient} magnetic field can be determined as
\begin{equation}\label{H_eff}
	\begin{array}{l}
	\displaystyle{H_{\text{eff}} = \left. \frac{1}{e} \frac{\partial \textbf{p}}{\partial t}\right|_\bot = 
	\left. \left(\textbf{E} + \frac{1}{c}\textbf{v} \times \textbf{B} \right)\right|_\bot = }\\
	\displaystyle{\sqrt{\left(\textbf{E} + \frac{1}{c}\textbf{v} \times \textbf{B}\right)^2 - \left(\frac{\textbf{p}}{\left| \textbf{p} \right|} \cdot \textbf{E}\right)^2}},
	\end{array}
\end{equation}
where $\textbf{p}$ is the particle's momentum, and $\textbf{E}$ and $\textbf{B}$ are the local electric and magnetic fields for the particle.

In the classical case the emission spectrum can then be determined by the expressions (\ref{cl_freq}) and (\ref{cl_spectrum}), and the emission is orientated along the direction of propagation with an angular spread of about $\gamma^{-1}$. However, as it is easy to understand, accounting for the synchrotron type of emission implies that the particle rotates through an angle of $\gamma^{-1}$ during a time interval shorter than the typical time of the EM-field variation. Indeed, the evident example of breaking this requirement is an electron passing through an undulator in the betatron regime resulting in a spectrum with $\gamma$ times narrower spread \RefA{compared to the synchrotron spectrum.} As it is easy to demonstrate (see \cite{wallin.pop.2015}), the betatron type of emission requires $\gamma$ times faster variation of field compared to the optical wave variation. In other words, assuming $\gamma \sim a \gtrsim 100$, it means that the individual particle emission remains the synchrotron type with the exception of specific cases, which imply transformation of an essential part of laser radiation energy into energy of higher than 100-th harmonic at least (in coherent form).

\subsection{Transition to quantum description}

The evident limitation of the classical expression for the synchrotron emission is that it implies the emission of photons whose energy can be larger than the electron has. In case of strong fields, this leads to overestimation of the spectrum spread and of the radiation losses rate. The simple estimate for the transition between classical and quantum description is commonly characterized by the dimensionless parameter $\chi$ defined by the Lorentz invariant expression \cite{LL.V4}:
\be\label{chi-def}
  \begin{array}{l}
	\displaystyle{\chi = \frac{e\hbar}{m^3c^4}\sqrt{p^\mu F^2_{\mu\nu} p^\nu} = }\\
	\displaystyle{\frac{e\hbar}{m^3c^4} \sqrt{\left(\frac{\epsilon\textbf{E}}{c} + \textbf{p}\times\textbf{H}\right)^2 - \left(\textbf{p}\cdot\textbf{E}\right)^2}.}
	\end{array}
\ee
As one can see from eq. (\ref{cl_freq}) and (\ref{H_eff}), the $\chi$ parameter has a simple meaning for the classical synchrotron theory; it determines the ratio of the typical photon energy to the electron kinetic energy:
\begin{equation}
  \chi = \frac{2}{3}\frac{\hbar \omega_s}{mc^2\gamma}
\end{equation}
(the factor 2/3 can be attributed to the definition of the typical photon energy). Thus, the values $\chi \ll 1$ correspond to the classical case, whereas $\chi \gtrsim 1$ indicates that quantum corrections are essential.

As one can see from the above-mentioned expressions, on the other hand the parameter $\chi$ represents a measure of transverse acceleration:
\begin{equation}\label{chi_H_eff}
  \chi = \gamma\frac{H_{\text{eff}}}{E_S},
\end{equation}
where $E_S = m^2c^3/e\hbar\simeq 10^{18}$ V/m is the Sauter-Schwinger limit. The second simple meaning of $\chi$ is ratio of the efficient magnetic field to $E_S$ in the rest frame of the particle.

Note that the classical expression for the total intensity of emission can be given via $\chi$ parameter:
\begin{equation}
  I^{cl} = \frac{2}{3}\frac{e^2 m^2 c^3}{\hbar^2} \chi^2.
\end{equation}
Assuming that the photons are emitted against the direction of propagation, we can determine the average force originated from recoils due to emission of photons:
\begin{equation}
  \textbf{f}_{RR}^{cl} = -\frac{2}{3}\frac{e^2 m^2 c}{\hbar^2} \chi^2 \textbf{v}.
\end{equation}
As one can see this expression coincides with the dominant (for ultra-relativistic case) term in the expression for the radiation reaction force in the Landau-Lifshitz form \cite{LL.V2}.

\subsection{Discreteness of radiation losses}
One consequence of quantum effects, in particular the quantization of emission, is the discreteness of radiation losses when e.g.~ an electron emits photons, as described in Sect.~\ref{NLC1} and Sect.~\ref{NLC2}. We can define a typical time interval between acts of photon emission as the ratio of the typical photon energy $\hbar \omega_t$ to the total radiation intensity $I$,
\be
  \tau_{t} = \frac{\hbar \omega_t}{I} \;.
\ee 
If $T_{t}$ is the typical time scale of the problem of interest, then we can characterize the discreteness of emission by the dimensionless parameter
\be
  \xi = 2\pi\frac{\tau_{t}}{T_{t}} \;.
\ee
If $\tau_{t}$ is small enough as compared with the time scale of the problem ($\xi \ll 1$), one can expect that discreteness of radiation losses can be smoothed out and thus can be reasonably well described by a continuous radiation reaction force. In fact, as it is well known (but is perhaps counterintuitive), the interval between photon emission is large ($\xi \gg 1$) for the non-relativistic problem of an electron rotating in a constant magnetic field $B$:
\be
	\xi = 2\pi\frac{\hbar c}{e^2} \frac {3}{2 \omega_B} \left(\frac{c}{v}\right)^2 \frac{\omega_B}{2\pi} \approx 200 \frac{c^2}{v^2},
\ee
where $v$ is the electron's velocity and $\omega_B$ is the frequency of rotation associated with both the typical photon energy ($\hbar \omega_B$) and the typical time scale ($2\pi/\omega_B$). Nevertheless, the effect of discreteness can hardly impact the classical results because the energy of photons is much less than the electron's energy ($\chi \ll 1$).

Using the classical expressions for synchrotron emission, we can demonstrate in the ultra-relativistic regime that an increase of $\gamma$ casues $\xi$ to decrease:
\be
	\xi^{cl} = 2\pi\frac{\hbar c}{e^2} \frac{9}{4} \frac{mc}{eB} \frac{\omega_B}{2\pi} \approx 300 \gamma^{-1}.
\ee


In the context of high intensity laser-matter interactions we can use the laser period $T_L$ as a typical time scale, and consider the laser amplitude as a typical effective magnetic field:
\be
  \tau_{t} = 2\pi\frac{\hbar c}{e^2} \frac{9}{4} \frac{mc}{eH} \approx \frac{\hbar c}{e^2} \frac{9}{4} \frac{T_L}{2\pi} a^{-1} \approx 300 T_L a^{-1},
\ee
giving
\be
  \xi^{cl} \approx 300 a^{-1}.
\ee
For the case of $\chi \gg 1$ we can use the approximate expression for intensity of emission in quantum regime \cite{LL.V4}:
\be\label{I_q}
  I^q \approx 0.37 \frac{e^2 m^2 c^3}{\hbar^2} \chi^{2/3},
\ee
and consider $mc^2\gamma/2$ as a typical photon energy:
\be
  \tau_{t} \approx \frac{\pi}{0.37}\frac{\hbar^2}{e^2mc}\gamma \chi^{-2/3}.
\ee
Assuming $\gamma \approx a$, we can obtain
\be
  \xi^q \approx 200 \left(\frac{\lambda}{\lambda_C} a\right)^{-1/3}
\ee
where $\lambda_C = 2\pi\hbar/mc$ is the Compton wave length, and $\lambda$ is the laser wave length.

Assuming $\gamma \approx a$, we can estimate the $\chi$ parameter in the context of laser-matter interaction problems:
\be
  \chi \approx a \gamma \frac{\omega_L \hbar}{m c^2} \approx a^2 \frac{\lambda_C}{\lambda}.
\ee

\begin{figure}
\centering\includegraphics[width=0.9\columnwidth]{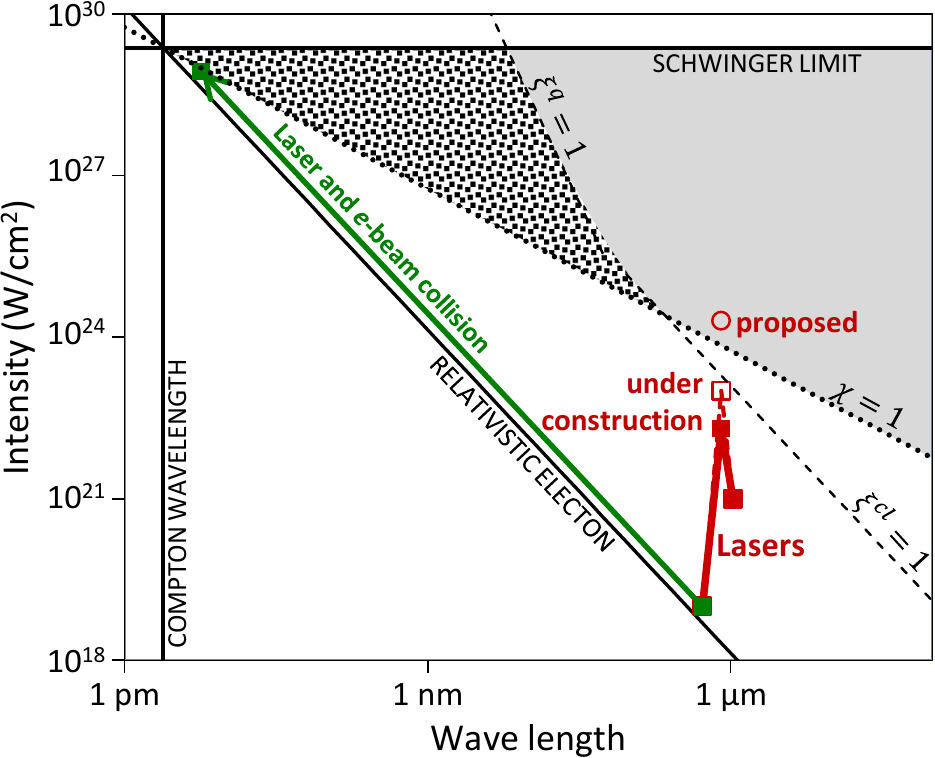}
\caption{(Color online) Regions of qualitatively different effect of electron individual emission on its motion. Fundamental values of Schwinger limit, Compton wave length and intensity threshold of relativistic motion of electrons are shown with solid lines. Some experimental capabilities are shown with red (ultraintense lasers) and green (laser pulse collision with external electron beam, SLAC experiment E144 \cite{burke.prl.1997}) lines.} \label{xichi}
\end{figure}

As we can see, in terms of quantum corrections to the classical description, with increase of laser intensity we have a competition of two counteracting effects: the time interval between photon emissions decreases, whereas the ratio of the photon energy to the electron energy increases. Summarizing this we can plot two curves on the plane of laser intensity and wave length (see Fig.~\ref{xichi}). The first one $\xi = 1$ (that corresponds to $2\pi$ events per period) qualitatively separates the parameter region into two parts: the recoil events are rare below and frequent above it. The second one $\chi = 1$ qualitatively separates parameter region into parts of weak recoils (below) and strong recoils (above). 

In terms of quantum corrections we have three qualitatively different regions. In the first region (white, $\chi < 1$) corrections are totally insignificant. In the second region (dotted, $\chi > 1$, $\xi > 1$) recoils are strong and infrequent, thus taking into account the quantum nature evidently can dramatically change results as compared with those obtained based on classical description. And the third region ($\chi > 1$, $\xi < 1$) is quite ambiguous. Every recoil is large but the recoils happen frequently. Thus classical description based on continuous force in some sense corresponds to "smoothed" recoil and may be considered as averaging of quantum description. Assuming (\ref{I_q}), for the approach of continuous force one should use a quantum expression for the radiation reaction force:
\begin{equation}
  \textbf{f}_{RR}^{q} \approx -0.37\frac{e^2 m^2 c}{\hbar^2} \chi^{2/3} \textbf{v}.
\end{equation} 

To understand the capabilities and limitations of the continuous force approach one can consider the problem of an electron moving in strong, constant, perpendicular electric and magnetic fields with $E>B$. The description based on the continuous force gives a saturation mode in which the gain in energy due to electric field is exactly compensated by the radiation losses. Thus the electron moves with some constant energy which corresponds to this saturation. More realistically, electrons in the quantum simulations emit photons instantly and between these events they are accelerated without action of the radiation reaction force. Averaging of this process roughly coincides with the description based on continuous force. The evident difference is the presence of a stochastic spread of electrons energy. The typical size of this spread is $\chi$ multiplied by the average electron energy. In particular, this implies that an electron can gain more energy than is predicted by the continuous force description. The difference in energy appears stochastically and has a non-zero probability even for very high values. 

\subsection{Spectrum of emission}

The classical theory of synchrotron emission does not take into account the fact that the energy of the emitted photon cannot exceed the energy of the electron. Thus, the classical rate of emission is not valid when the typical photon energy $\hbar \omega_s$ is compatible with the electron energy $m c^2 \gamma$. (This happens at intensities order of $10^{24}$~W/cm$^2$ and higher.) As a measure of radiative losses we therefore define the dimensionless parameter $\delta = \hbar \omega / \left(mc^2 \gamma\right)$, which is just the emitted photon energy relative to the emitting electron energy.

The classical expression for the spectral power of synchrotron emission (valid for $\delta \ll 1$), assuming a constant, crossed background field, can be written \cite{LL.V2}
\be\label{Iw_cl}
\begin{array}{l}
  \displaystyle{\frac{\partial I^{cl}}{\partial \omega} = \frac{\sqrt{3}}{2\pi} \frac{e^3 H_{\text{eff}}}{mc^2} F_1\left(z_{cl} \right),}\qquad \displaystyle{z_{cl} = \frac{2}{3} \chi^{-1} \delta,} \\
\end{array}
\ee
where $F_1(x) = x \int_x^\infty dt K_{5/3}(t)$ is the first synchrotron function. The corresponding expression valid for high energy emissions ($\delta \sim 1$), i.e.\ that derived from QED, is given in~\cite{nikishov.jetp.1964, nikishov.jetp.1967, baier.pla.1967, LL.V4}. In the way similar to derivation of \ref{Iw_cl} (see \cite{LL.V2}), we represent this expression via the first and the second synchrotron function $F_2(x) = x K_{2/3}(x)$:
\be\label{Iw_q}
\begin{array}{l}
  \displaystyle{\frac{\partial I^{q}}{\partial \omega} = \frac{\sqrt{3}}{2\pi} \frac{e^3 H_{\text{eff}}}{mc^2} \left(1 - \delta\right) \left\{
	F_1\left(z_{q} \right) + \frac{3}{2} \delta \chi z_q F_2\left(z_q\right)\right\},} \\
	\displaystyle{z_{q} = \frac{2}{3} \chi^{-1} \frac{\delta}{1 - \delta}.}
\end{array}
\ee 
This form is convenient for further analysis and numerical computation. In particular, as one can see the quantum expression (\ref{Iw_q}) fits the classical one (\ref{Iw_cl}) in the limit of $\delta \ll 1$. The rate of photon emission can be obtained as the spectral power divided by photon energy $\hbar \omega$.

\subsection{Event generator}
In the regime $\chi \gtrsim 1$ quantum effects become important and it is not possible to interpret physical processes in purely classical terms; this was already touched upon in Sect.~\ref{SECT:HIGHER}. The method applied in the PIC approach, which attempts to account for this discrepancy, is to assume that events of e.g.\ photon emission occur stochastically and instantaneously according to probability distributions calculated in QED, and that between these events the particles'€™ motion can be described classically (as e.g.\ motion under the Lorentz force)~\cite{nikishov.jetp.1967, baier.jetp.1968, ridgers.jcp.2014}. As already discussed, recoil accompanies each emission event. In the most general case, this \RefA{model implies adding} to the traditional PIC approach an \textit{event generator} based on the probability distribution (\ref{Iw_q}).

In terms of computing, our problem is the development of a numerical algorithm that uses a random number generator to make (at given time steps) a decision of photon emission, and to choose an energy for that photon, in such a way that the distribution function of emitted photons coincides with theoretical distribution (\ref{Iw_q}). Here we focus on issues specific to the implementation of the event generator in studies of ultra-strong laser-matter interactions.

To make the discussion self-sufficient we start from the implementations (sometimes referred as \textit{QED-MC (Monte-Carlo)} or \textit{QED-PIC}) that have been recently proposed and developed for this purpose \cite{elkina.prstab.2011, sokolov.pop.2011, duclous.ppcf.2011, ridgers.jcp.2014}. 

The most common implementation for the event generator implies the following procedure at each time step $\Delta t$ for each particle. First, using particles position and momentum, the $\chi$ parameter is calculated according to (\ref{chi-def}) (see e.g.~\cite{wallin.pop.2015} for details). Using this, the {\it total} probability $P$ of emitting a photon, during the current time step, is calculated by integrating (\ref{Iw_q}) over all possible $\delta$ (see \RefA{(\ref{W_generator}), and the text following below, for the explicit expression).} Next, we generate a random number $r_1\in[0,1]$.  If $r_1 < P$ then we say a photon has been emitted, otherwise we do nothing. If a photon is emitted, we must determine its energy and direction. Energy is determined based on inverse transform sampling. A second random $r_2\in[0,1]$ is generated, and the photon energy $\delta$ is computed as the root of the equation
\be
	\frac{W(\delta)}{ W(1)} = r_2,
\ee
in which $W(x)$ is the probability for the electron to emit a photon within the range of energies from $\delta_{\text{min}} mc^2\gamma$ (the small low-energy cutoff $\delta_{\text{min}}$ is used to avoid the integrable singularity at $\delta=0$) and $x mc^2\gamma$, during the interval of time $\Delta t$:
\be\label{W_generator}
	W(x) = \Delta t \int\limits_{\delta_{\text{min}}}^{x}\!\ud\delta\  \frac{\partial I^{q}}{\partial \omega} \frac{1}{\hbar \omega}.
\ee
The total probability $P$ is simply $P = W(1)$. The photon's direction of propagation (wave vector) is commonly chosen along the electron's momentum vector because, as mentioned above, the typical spread about this direction is $\mathcal{O}(1/\gamma) \ll 1$ in the ultra-relativistic case. Once this is done, we add a new photon with the obtained properties and reduce the momentum of the parent particle to satisfy conservation of energy and momentum. In terms of the PIC formalism, the procedure is applied to each super-particle, which is associated with some number of real particles having the same properties. Thus, the event of emission may be implemented as the creation of a super-photon, which is associated with the same number of real photons. Some other possibilities are described in Sect.~\ref{controlling_computational_expenses}.

Implementation of the event generator as described is rather straightforward. However, it has a number of undesirable features:
\begin{enumerate}
	\item The integration (\ref{W_generator}) is computationally expensive.
	\item Low energy photons with $\hbar \omega < \delta_{\text{min}} m c^2 \gamma$, cannot be included because of the infra-red cutoff $\delta_{\text{min}}$.
	\item Since no more than one photon can be generated within each time step, the requirement $P\ll 1$ imposes an additional limitation on the time step.
\end{enumerate}
We will therefore now describe some ways of overcoming these difficulties.

To avoid the integration (\ref{W_generator}) at each occasion, one can in advance tabulate, and store in memory, $W$ for some array of possible values of $\chi$, and use interpolation during the simulation~\cite{ridgers.jcp.2014}. Alternatively, an efficient \textit{alternative event generator} has been proposed in~\cite{elkina.prstab.2011}. This method requires generation of two random variables $\{r_1,r_2\}\in[0,1]$ with uniform probability. Next, if $r_2 < P\left(r_1\right)$ we assume emission of a photon with energy $r_1 mc^2\gamma$, and do nothing otherwise. Here, $P\left(\delta\right)$ is the probability density for emission of photon with energy $\delta mc^2\gamma$ defined as
\be\label{P_q}
\begin{array}{l}
\displaystyle{P\left(\delta\right) = \frac{\partial I_{q}}{\partial \omega} \frac{\partial \omega}{\partial \delta} \frac{1}{\hbar \omega} \Delta t =} \\
\displaystyle{\left[\Delta t \frac{e^2mc}{\hbar^2}\right] \frac{\sqrt{3}}{2\pi}\frac{\chi}{\gamma} \frac{1 - \delta}{\delta}\left\{ F_1\left(z_{q} \right) + \frac{3}{2} \delta \chi z_q F_2\left(z_q\right) \right\}}.
\end{array}
\ee
If $P\left(\delta\right) < 1$ for all values of $\delta \in \left[0, 1\right]$, this expression in average leads to photon distribution in accordance with the spectral power (\ref{Iw_q}). 

However, $P\left(\delta\right) \approx \left(\Delta t e^2mc/\hbar^2\right) \times 0.52 \gamma^{-1} \chi^{2/3} \delta ^{-2/3}$ in the limit of $\delta \ll 1$, and thus is singular in the point $\delta = 0$. Therefore for any value of the time step there is a vicinity of zero point where the requirement $P\left(\delta\right) < 1$ fails. Thus, this method also includes the problem of regularization. As one can understand, instead of the complete cutoff the alternative event generator underestimates the rate of emission for the photons of some low energy region ($P(\delta) > 1$) determined by the time step (see Fig.~\ref{generators}).  Fortunately, for simulation of cascading processes \cite{nerush.prl.2011} there is a natural cutoff: photons of energy $\hbar \omega < 2mc^2$ are unlikly to produce pairs in $E < E_S$. As a result, the physics of cascades can be captured for an appropriate choice of $\delta_{\text{min}}$ or time step. (The underestimation of radiation losses in this case is negligible, having a relative factor of less than $1/\gamma$.)

\subsection{Modified event generator}

However, in the interests of accurate simulation, we should ask if we can do better than to simply ``add a cutoff"; in particular, can we construct a simulation which produces sensible diagnostics for all photon energies. To avoid the problems associated with emission of low energy photons we propose the following simple modification of the alternative event generator. This is designed to produce more photons at low frequency, so that the photon emission spectrum agrees well with the theoretical result for the whole range of frequencies.

We again generate two random variables $\{r_1, r_2\}$ from the interval $[0,1]$ with uniform probability. But now we assume a photon is emitted with energy $r_1 mc^2\gamma$ only if $r_2 < P_m\left(r_1\right)$ where
\be
	P_m(r_1) := \frac{\partial f_m\left(r_1\right)}{\partial r_1} P\big(f_m(r_1)\big) \;,
\ee
in which $f_m(r_1) $ is a function chosen to cancel the singular behaviour of the probability function $P$ in the low energy limit. (If $r_2>P_m(r_1)$ we do nothing.) We call this the ``\textit{modified event generator}''. One can show that $f_m(x) = x^3$ solves the problem and, for an appropriately chosen time step (which ensures that $P_m(r_1) < 1$ for $r_1 \in [0, 1]$) the resulting spectrum agrees perfectly with the analytic expression.
 
\begin{figure}[t]
\centering\includegraphics[width=1.0\columnwidth]{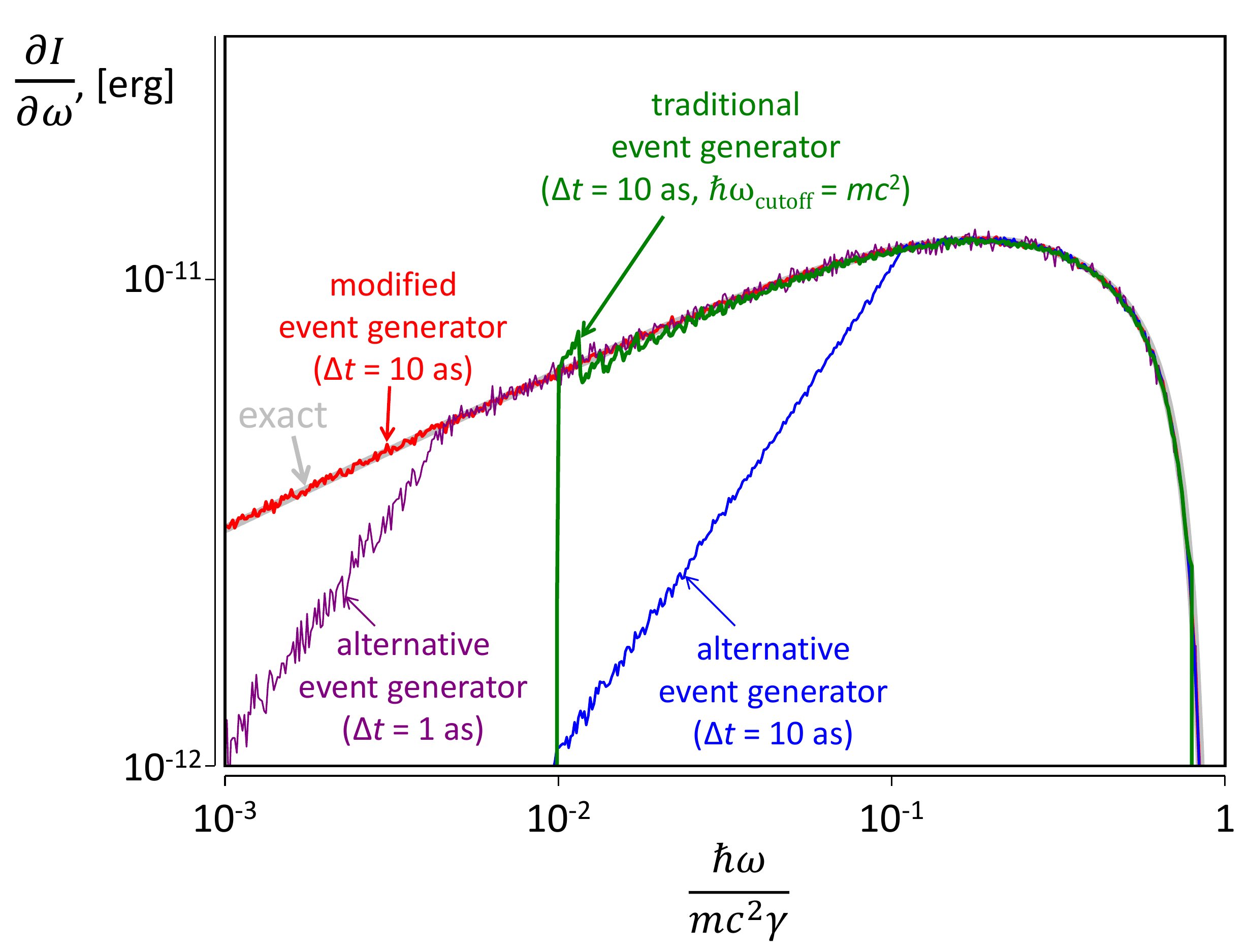}
\caption{(Color online) The emission spectrum for $\gamma = 100$, $\chi = 1$, obtained from the exact expression (\ref{Iw_q}) (gray) and from running various event generators; the traditional event generator with a cutoff (green), the alternative event generator (blue and purple), and our modified event generator (red). Clearly the modified event generator gives excellent agreement with the analytic result. \RefB{The absence of a low-energy cutoff allows us to model the entire range of spectral emission.}
}
\label{generators}
\end{figure}

The three considered event generators are compared in Fig.~\ref{generators}, in which we plot the emission spectrum of an electron obtained from running different event generators 10$^7$ times for each value of photon energy, keeping fixed the parameters $\gamma = 100$ and $\chi = 1$. Note, that taking into account low energy photons makes the modified algorithm capable of correct simulation of the radiation reaction force in the low energy limit. There is therefore no need to merge this method with an additional method which accounts for radiation reaction classically -- in fact, doing so typically raises the problem of double counting. 

\subsection{Adaptive event generator}\label{SECTION:AEG}

The requirement of $P_m\left(r_1\right) < 1$ provides correct energy distribution of generated photons. Nevertheless, if $P_m\left(r_1\right) \sim 1$ the photons are generated every iteration with relatively high probability. As a result almost equidistant emission of photons appears as an artificial numerical effect. This can affect results, for example, in case of electron acceleration balanced by radiation reaction \cite{gonoskov.prl.2014}. To avoid this effect we should use a stronger requirement: 
\be
P_m\left(r_1\right) \ll 1.
\ee
The expression for $P_m\left(r_1\right)$ is combined with two terms that essentially determined by the first and the second synchrotron function. The first one has its maximum in the zero point, which we can obtain using asymptotics of the first synchrotron function ($F_1\left(x \ll 1\right) \approx 2.15 x^{1/3}$). The second term has its maximum in the vicinity on unity (for high enough intensities), which is determined by the global maximum of the second synchrotron function (max$\left(F_2\left(x\right)\right) \approx 0.6$). In such a way, for the modified event generator we can obtain two requirements that correspond to the first and the second term respectively:
\be
\begin{array}{l}
 \displaystyle{\left[\Delta t \frac{e^2mc}{\hbar^2}\right] \times 1.5 \chi^{2/3} \gamma^{-1} \ll 1,}\\
 \displaystyle{\left[\Delta t \frac{e^2mc}{\hbar^2}\right] \times 0.5 \chi \gamma^{-1} \ll 1.}
\end{array}
\ee
The requirement for the time step can be written in the form:
\be
\Delta t \ll \frac{137}{2\pi}\frac{\lambda_C}{c} \min\left\{ 0.67 \gamma^{1/3} \left(\frac{E_s}{H_{\text{eff}}}\right)^{2/3}, 2\left(\frac{E_s}{H_{\text{eff}}}\right) \right\}
\ee
or, assuming $\gamma \geq 1$ and $E_s/H_{\text{eff}} \geq 1$, in a simplified form: 
\be\label{dt_req}
	\Delta t \ll 14 \frac{\lambda_C}{c} \left(\frac{E_s}{H_{\text{eff}}}\right)^{2/3}.
\ee

As the intensity grows this requirement becomes stronger and for high intensities can be stronger than all other requirements of the PIC approach. One can just reduce the time step to follow this requirement. Nevertheless this can evidently cause reduction of the computational efficiency. To overcome this problem we propose an \textit{adaptive event generator}, which implies the following.

For each iteration, apart from the procedures of the PIC approach for the standard time step $\Delta t$ (without requirement (\ref{dt_req})), we run a loop over all particles to account for particles emission. For each particle we calculate an individual time step $\Delta t_q$ that fulfills the requirement (\ref{dt_req}). If $\Delta t_q < \Delta t$ we divide the time step $\Delta t$ into subintervals not greater than $\Delta t_q$ and run a loop, which includes the particle pusher and the event generator. We assume here that EM-field remains almost the same during $\Delta t$, which is guaranteed by the standard requirements for the time step of the PIC approach. This {\it subdivision} method provides optimal computational loading by using individual time step values for each particle and each instant of time. Moreover it does not require more frequent interpolation of EM-field, which is usually the most computationally expensive operation. The implementation of this algorithm is tested and benchmarked in Sect.~\ref{SECT:CASCADE}.

\section{Pair production}

\subsection{Stimulated pair production}
As introduced above, stimulated pair production describes the creation of an electron-positron pair from a hard photon interacting with a strong EM-field. As the Feynman diagram for this process is related to that of nonlinear Compton scattering by crossing symmetry, it is not surprising that the probability density for pair production may be written in a form similar to (\ref{P_q})~\cite{LL.V4}: 
\be\label{P_p}
\begin{array}{l}
		\displaystyle{P_{\text{p}}\left(\delta_e\right) = \left[\Delta t \frac{e^2mc}{\hbar^2}\right] \frac{\sqrt{3}}{2\pi} \times}\\
		\displaystyle{\frac{\chi_{\gamma}}{\hbar\omega/\left(mc^2\right)} \left(\delta_e - 1\right)\delta_e\left\{ F_1\left(z_{\text{p}} \right) - \frac{3}{2} \chi_{\gamma} z_{\text{p}} F_2\left(z_{\text{p}}\right) \right\}}, \\
		\displaystyle{z_{\text{p}} = \frac{2}{3}\frac{1}{\chi_{\gamma} \left(1 - \delta_e\right)\delta_e},}
\end{array}
\ee
where $\hbar \omega$ is the photon energy, whereas the energies of electron and positron are denoted as $\delta_e \hbar \omega$ and $\left(1 - \delta_e\right) \hbar \omega$ respectively. The EM-field enters the probability via the parameter $\chi_{\gamma}$ defined as
\be
\chi_{\gamma} = \frac{\hbar \omega}{mc^2} \frac{H^{\gamma}_{\text{eff}}}{E_s}.
\ee
Here $H^{\gamma}_{\text{eff}}$ is again the effective magnetic field, defined in the same way as for particles (\ref{H_eff}), but relative to the direction of photon propagation:

\begin{equation}\label{H_gamma_eff}
	\begin{array}{l}
	\displaystyle{H^{\gamma}_{\text{eff}} =  \left. \left(\textbf{E} + \frac{c}{\omega}\textbf{k} \times \textbf{B} \right)\right|_\bot = }\\
	\displaystyle{\sqrt{\left(\textbf{E} + \frac{c}{\omega}\textbf{k} \times \textbf{B}\right)^2 - \left(\frac{\textbf{k}}{\left| \textbf{k} \right|} \cdot \textbf{E}\right)^2}},
	\end{array}
\end{equation}
where $\textbf{k}$ is the photon wave vector.

In contrast to the case of nonlinear Compton scattering, the expression (\ref{P_p}) is infra-red finite, and thus simulations of pair production can be implemented straightforwardly, based on the following event generator~\cite{elkina.prstab.2011}, without any additional numerical tricks.

At each time step and for each photon we generate two random values $\{r_1, r_2\}$, with uniform probability, in the interval $[0, 1]$. If $r_2 < P_{\text{p}}\left(r_1\right)$ we assume creation of an electron and positron moving in the same direction as the parental photon and with momentum of $r_1 \hbar \omega/c$ and $\left(1 - r_1\right) \hbar \omega/c$ respectively. In the PIC simulation itself, both the electron and the positron are added to the ensemble of particles as super-particles, associated with the same number of real particles as the parent super-photon was associated with: the parent super-photon is removed from the simulation. 

Similar to derivation of (\ref{dt_req}), one can obtain the requirement of $P_{\text{p}}\left( \delta_e \right) \ll 1$ from analysis of the expression (\ref{P_p}):
\be\label{dt_p_req}
	\Delta t \ll 130 \frac{\lambda_C}{c} \left(\frac{E_s}{H^{\gamma}_{\text{eff}}}\right).
\ee

\subsection{Non-perturbative pair production}

Pairs can be produced from the fields of the laser themselves, without the presence of seed particles. This process, originally considered in~\cite{Sauter:1931zz,Heisenberg:1935qt,Schwinger:1951nm} is typically nonperturbative in nature, and the rate of pair production is exponentially suppressed below electric fields of strength
\be
	E_S := \frac{m^2c^3}{e\hbar} \simeq 1.3\times 10^{18} \text{ V/m} \;.
\ee
This nonperturbative process can be taken into account using the following statistical routine. The electromagnetic field may be assumed constant over the size of a cell, the dimensions of which will be very much smaller than the laser wave length. Let the field strength (for simplicity of presentation neglecting the magnetic field) in the cell be $E$, constant. The rate $R_S$ at which pairs are created from the vacuum\footnote{Note that $R$, the pair production rate, should not be confused with the imaginary part of the effective action, which is associated with the rate of vacuum decay; a clear explanation of the difference is given in~\cite{Cohen:2008wz}.}, per unit four-volume, is given in QED by~\cite{Nikishov1970346};
\be
	R_S = \frac{1}{4\pi^3 \lambdabar_c^4}  \frac{E^2}{E_S^2} \exp\bigg[-\frac{\pi E_S}{E}\bigg] \;,
\ee
where $\lambdabar_c = \hbar/mc$ is the reduced Compton wave length. For each iteration and each cell of a grid we calculate the probability of pair production directly from the field as
\be
	P_S = R_S\times V_{\text{cell}} \times c\Delta t.
\ee
Next we generate a random value $r\in [0,1]$. If $r < P$ we produce an electron and a positron with zero momentum in arbitrary point of the cell, otherwise we do nothing. As observed above, the energy required to make the pair, $2mc^2$, is much lower than the energy lost from the intense field in accelerating the particles up to relativistic speeds. Hence, the energy loss of $2mc^2$ from the field is considered negligible in the code.

\section{Controlling computational costs}\label{controlling_computational_expenses}
Many studies have found that cascades can cause large increases in the number of particles and photons in a system. In terms of computational costs, this will raise both the memory required for super-particles/photons allocation, and the computational time required to handle them. The traditional PIC approach solves this problem by an appropriate choice of statistical weights which controls the number of super-particles representing the whole number of real particles. Traditionally, the choice of statistical weights is however made once, in the beginning of simulation; this implicitly assumes that the number of particles will remain constant, or change insignificantly. The evident solution for the problem of an increasing number of particles is to vary the statistical weights dynamically during the simulation. The efficient implementation of this idea requires two modifications:
\begin{enumerate}
	\item generation of new super-particles/photons should be performed with a tunable statistical weight,
	\item initial or previously generated super-particles should be replaced with those having the larger (tunable) statistical weight.
\end{enumerate}
The first modification is straightforward. We can define a tunable factor $f_g > 1$ for an event generator (of any nature), which implies the following. We execute the event generator with probability $1/ f_g$ or, alternatively, once every $f_g$ occasions, and multiply the statistical weight of the produced particles by $f_g$. (Note, that one can use a similar modification `in reverse' to reduce the statistical noise by running the event generator $N_g > 1$ times every occasion and multiplying the statistical weight of the produced particles by $1/N_g$.) \\

The second modification is more ambiguous. There are two well-known concepts: thinning and merging. The thinning concept \cite{timokhin.mnras.2010, nerush.prl.2011, lapenta.jcp.1994} implies removing random super-particle/photon from the simulation and distributing its statistical weight $f_\text{r}$ between either neighboring or all the remaining super-particles/photons of this type (for example, global even redistribution \cite{timokhin.mnras.2010} assumes multiplying particles statistical weights by $\sum f / \left(\sum f - f_\text{r}\right)$ ). The procedure fulfills charge conservation, but can cause a small local deviation from energy and momentum conservation. The concept of merging implies choosing a number of super-particles/photons (of the same type) located closely in the coordinate-momentum space and merging them into one or several super-particle/photon \cite{lapenta.jcp.2002, vranic.arxiv.2014}. This concept is certainly more computationally expensive and difficult for implementation, but it provides less numerical noise than the first one. The numerical scheme developed for the current study incorporates a number of simple implementations of the thinning concept.

\section{Numerical tests}\label{SECT:CASCADE}
We conclude this study by testing our implementation of the adaptive event generator (which is based on the modified event generator), \RefA{using some published results as benchmarks.} The implementation also incorporates a thinning procedure, which controls computational costs by restricting the number of super-particles located in a domain/cell or generated from a single particle/photon during one iteration.

\subsection{Cascade development}
\begin{figure}[h!!!]
\centering\includegraphics[width=1.0\columnwidth]{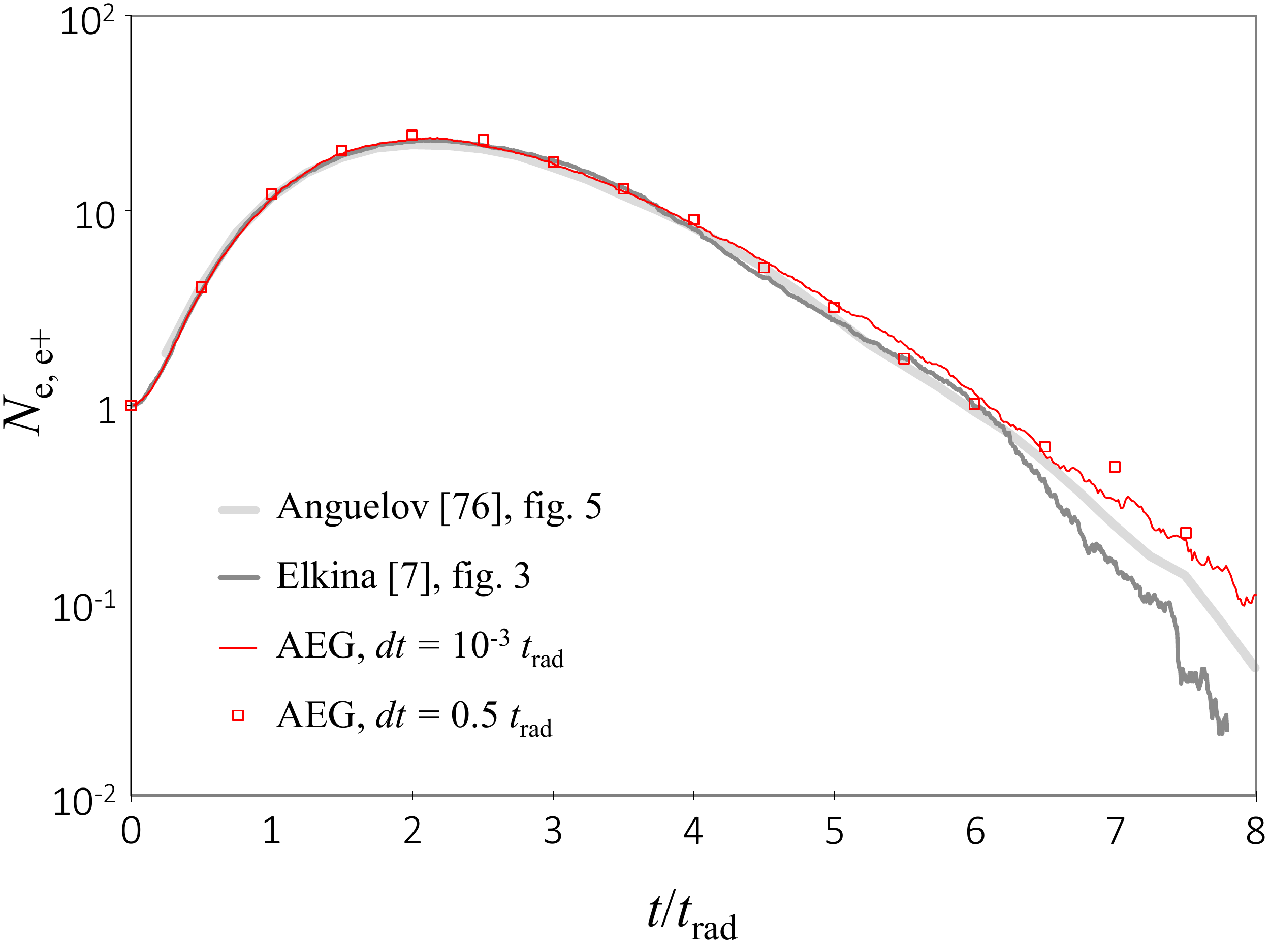}
\caption{(Color online) 
{\bf Cascade development:} Comparison of the number of particles produced in a cascade, under the conditions of the first test problem described in the text and in~\cite{anguelov.jpg.1999, elkina.prstab.2011}. The plot shows the number of produced particles with energy exceeding 0.1\% of the initial electron's energy, as a function of time. We show the results of~\cite{anguelov.jpg.1999} (gray heavy curve), \cite{elkina.prstab.2011} (dark gray curve) and of our implementation for the adaptive event generator (AEG) with small (thin red curve) and large (red squares) time steps. \RefA{As one can see, for this benchmark we find excellent agreement of our results with the published ones.} 
}
\label{anguelov}
\end{figure}
We consider first a test problem described in~\cite{anguelov.jpg.1999} and later used as benchmark in~\cite{elkina.prstab.2011}. The test problem assumes the development of a cascade from a single electron, initial gamma factor $\gamma_0 = 2 \times 10^5$, moving in a constant magnetic field of strength $H_0 = 0.2 E_S$ orientated perpendicularly to the electron's direction of motion. In Fig.~\ref{anguelov} we plot the number of produced particles with energy exceeding 0.1\% of the initial electron's energy, as a function of time. The plot shows our own results, as well as those of~\cite{anguelov.jpg.1999, elkina.prstab.2011}. Following~\cite{anguelov.jpg.1999, elkina.prstab.2011}, the parameter $t_{rad} = 3.85 \times \gamma_0^{1/3} \left(E_S/H_0\right)^{2/3} \hbar^2/\left( m c e^2 \right)$ is used as a typical time of radiation.  Our result, like that of~\cite{elkina.prstab.2011}, is obtained from averaging over $10^3$ simulation runs. As one can see, our results agree very well with those of~\cite{anguelov.jpg.1999, elkina.prstab.2011} for both small and large time steps. This confirms the validity of the time step subdivision performed in the AEG method.

\subsection{Distribution functions}
\begin{figure}[h!!]
\centering\includegraphics[width=1.0\columnwidth]{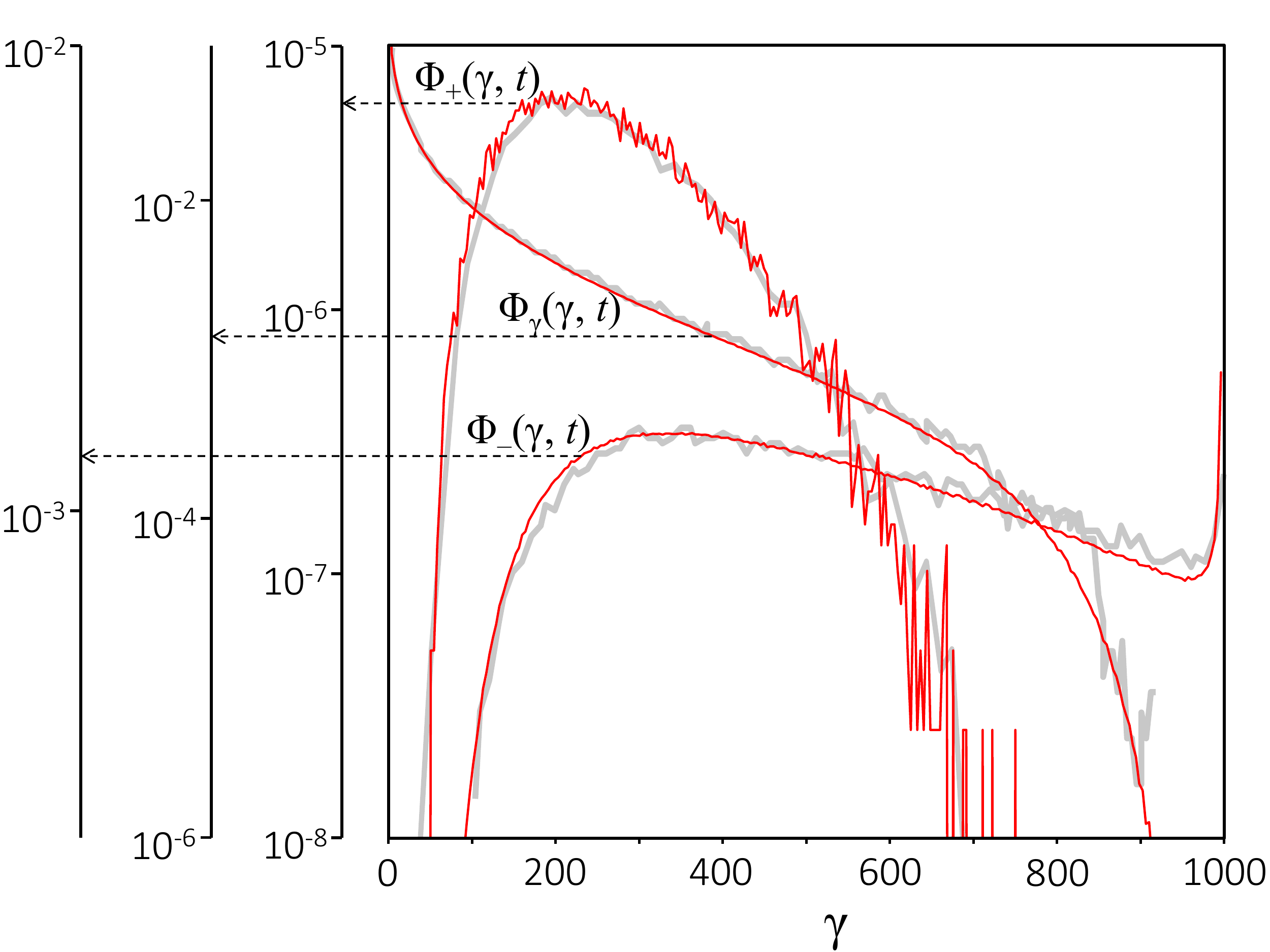}
\caption{(Color online) {\bf Distribution functions:} Particle distribution functions: comparison of our AEG implementation (thin red curves) with the results of \cite{ridgers.jcp.2014} (gray heavy curves).  \RefA{Excellent agreement is clearly seen for this benchmark.}
}
\label{ridgers}
\end{figure}
Next, we consider test problem number 1 from~\cite{ridgers.jcp.2014}, in which the authors benchmarked a numerical simulation against a direct numerical computation of analytical expressions for the evolution of the distribution functions. The problem assumes the same setup as above but with $\gamma_0 = 1000$ and $H_0 = 10^{-3} E_S$. In Fig.~\ref{ridgers} we plot the number of electrons, $\Phi_{-}(\gamma, t)$, photons $\Phi_{\gamma}(\gamma, t)$ and positrons $\Phi_{+}(\gamma, t)$ as a function of energy in units of $mc^2$, after 1~fs of the cascade development. Our results are averaged over 10$^7$ simulation runs. Comparing with the corresponding figures~2~(a), (c) and (e) from~\cite{ridgers.jcp.2014}, one can see that the results agree very well.

\subsection{Development of an avalanche-type cascade}
\begin{figure}[h!!]
\centering\includegraphics[width=1.0\columnwidth]{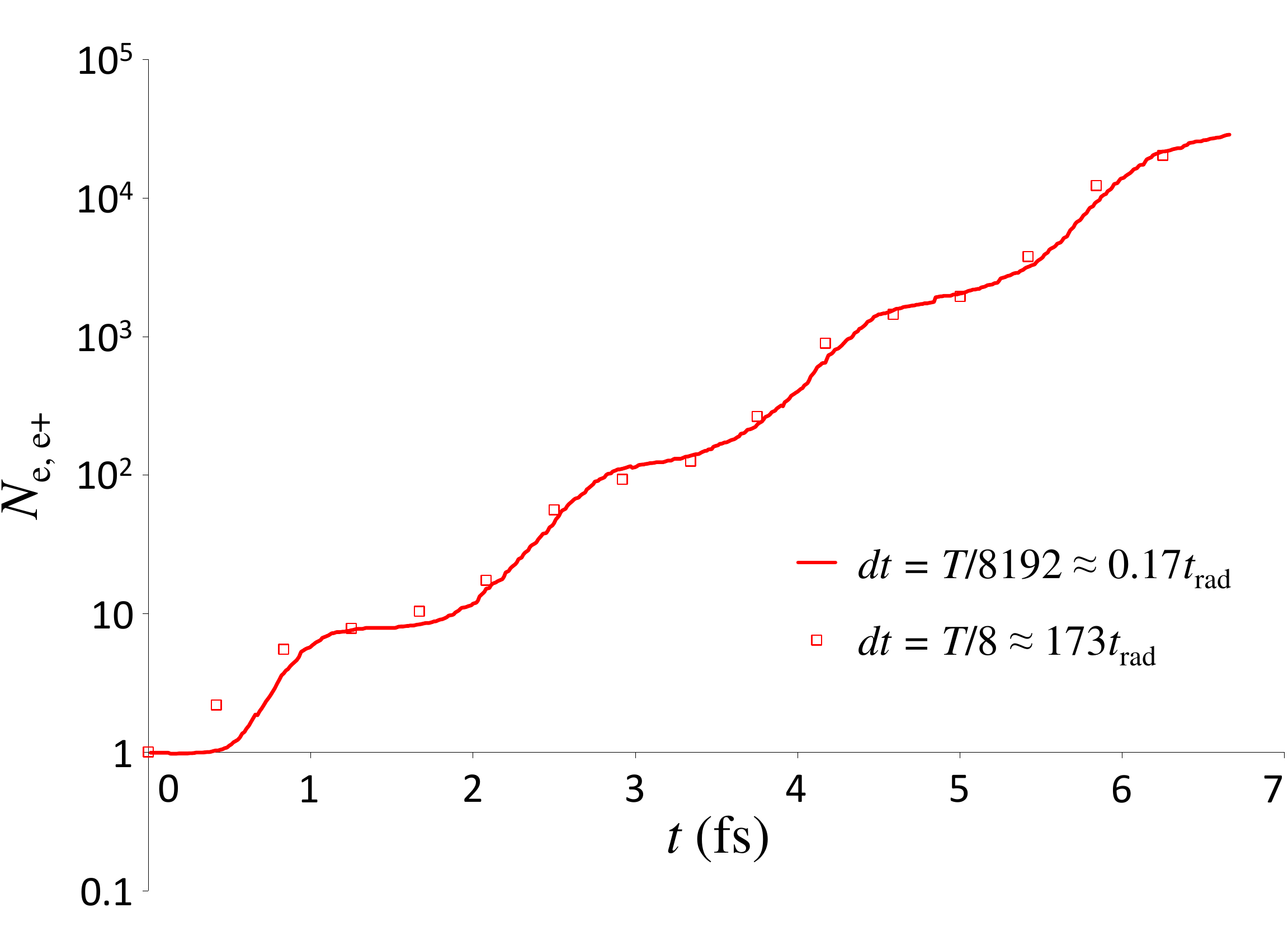}
\caption{(Color online) {\bf Avalanche-type cascade:} Number of particles produced in an avalanche-type cascade, as a function of time. The avalanche is initiated by an electron ($x = \lambda/8$ at $t = 0$) with $p_y = 1000 mc$ in a plane standing wave defined in the form $E_y (t = 0) = 10^{-2} E_S \sin(2\pi x/\lambda)$ ($\lambda = 1 \mu$m). The results are obtained from an AEG implementation with $dt = 0.17 t_{rad}$ (thin red curve) and $dt = 173 t_{rad}$ (red squares). 
\RefA{The agreement between these results indicates that the time step and particles/photons subdivision procedures allow much more efficient use of computational resources without affecting the physical results.}
}
\label{ACascade}
\end{figure}
The two test problems above are sufficient for testing the basic rates of photon emission and pair production. However, they are not sensitive to particle motion and do not require active use of thinning procedures. In addition, the produced particles/photons do not gain energy from the field, but essentially share the energy of the initial electron. This rather refined process has been called a \textit{shower-type cascade} in~\cite{mironov.arxiv.2014}.

To test our numerical scheme in more sophisticated conditions we consider now an \textit{avalanche-type cascade} \cite{mironov.arxiv.2014}, in which particles experience rapid energy gain from a strong field. (The subsequent emission of hard radiation leads to pair creation and seeds the cascade.) We consider an electron having initial gamma factor $\gamma_0 = 1000$ as a seed particle. The cascade is induced by a plane standing wave. At the initial instant the EM-field has only a single component $E_y (t = 0) = 10^{-2} E_S \sin(2\pi x/\lambda)$, and the electron moves along the $y$-axis at $x = \lambda/8$, where $\lambda = 1~\mu$m. (In this regime, we remark that trapping effects become important -- see for example \cite{kirk.ppcf.2009, gonoskov.prl.2014}.)

In Fig.~\ref{ACascade} we plot the number of produced electrons and positrons obtained from simulations as a function of time, with two different values of time step. Using the small value $dt = 0.17 t_{rad}$ does not lead to subdivision of the time step, while for the large step value $dt = 173 t_{rad}$ the subdivision is performed at almost every occasion (here $t_{rad} = 14 \left(E_S/H_0\right)^{2/3}\lambda_C/c$ in accordance with \ref{dt_req}). The close agreement between the results for two different time steps shows that the subdivision does not greatly impact the results\RefA{, while drastically reducing computational time}. (Certainly, the results are slightly affected due to the fact that for $dt = 173 t_{rad}$ we update the field only 8 times per cycle.) In addition, we performed the simulation with and without thinning procedures, and varied their parameters, confirming that the result is not affected by these procedures. \\

\RefA{We have therefore demonstrated that the proposed adaptive event generator allows to account for multiple QED events within each iteration by performing optimal and appropriate subdivision of the time step of a PIC simulation. In such a way, we uncouple the time-step constraints originating in the PIC scheme from those originating in the statistical routine. This makes computations significantly more efficient without losing accuracy. This can of course drastically decrease the computational time if ultra-strong fields are reached in a focal region that occupies a small part of the simulation box.}

\section{Conclusions}\label{SECTION:CONCLUSIONS}


Particle-in-cell schemes today constitute a standard tool in the numerical modeling of plasmas, and in particular laser produced plasmas. It was realised early on that one of the challenges of such codes would be to include physics of different scales in a single run \cite{dawson.rmp.1983}. Thus, the development of PIC schemes to include high-energy processes is part of a long-standing problem. The approach presented in this paper cannot be said to fully address all aspects of such multi-scale issues, but is rather a pragmatic way to push the codes further in their domain of applicability. In this pragmatic sense, we have described the extension of traditional PIC approaches required for studying physics in strong laser fields. We have described several methodological and algorithmic problems which arise in this extension, and presented possible solutions. \RefA{In particular, we have shown how to avoid a low-energy cutoff in the incoherent emission spectrum, and how to uncouple time-step constraints originating in the PIC scheme from  those originating in the statistical routines.}

\RefA{We have benchmarked our implementation of our adaptive event generator against results in the literature. The proposed ideas and the codes based on them, such as that developed here, will be a useful tool in many upcoming experimental campaigns, and will ideally assist in the development of novel experiments where high-intensity physics meets high-energy physics~\cite{Dunne:2008kc,Heinzl:2009zd,Jaeckel:2010ni,Redondo:2010dp,Villalba-Chavez:2013bda}.}

\section{Acknowledgments}

The research is supported by the Ministry of education and science of the Russian Federation (the agreement of August 27, 2013 No. 02.B.49.21.0003 between The Ministry of education and science of the Russian Federation and Lobachevsky State University of Nizhni Novgorod), by the Russian Foundation for Basic Research (Projects 14-07-31211, 14-02-31495 and 15-37-21015), by the Swedish Research Council (Grants 2010-3727, 2011-4221 and 2012-3320) and by the Wallenberg Foundation grant `{\it Plasma based compact ion sources}' . The simulations were performed on resources provided by the Swedish National Infrastructure for Computing (SNIC) and the Joint Supercomputer Center of RAS.

\section{Appendix}
\subsection{Interface for merging modules}\label{SECTION:MERGE}

The modifications described above do not interfere with the implementation of standard PIC processes. Thus, these modifications may be included in the form of independent add-ons, or modules, that do not rely on any specific PIC implementation. This approach has a number of advantages. First, it becomes possible to test and use the same modules with different PIC implementations. Second, abstracting from the details of the PIC implementation and parallelization simplifies the process of module development. Third, development of PIC implementations and modules can be performed simultaneously. Here we present an interface, called a module development kit for PIC approach (PIC-MDK), which is intended to provide a sufficient link between an arbitrary module and an arbitrary PIC implementation. We note that despite the mentioned advantages the interface is designed in the way that it does not slow down any of the performance-critical parts. \\

PIC-MDK assumes C++ object-oriented programming. Each module is supposed to be implemented in the form of a separate class inherited from a base class \texttt{Module}. A number of virtual functions of the \texttt{Module} class can be overloaded to implement handling various unified occasions for the modules, such as initialization, saving and loading. We suppose parallel simulations, thus one object of class \texttt{Module} for each enabled module is created for each domain of the grid.\\

\begin{figure}
\centering\includegraphics[width=1.0\columnwidth]{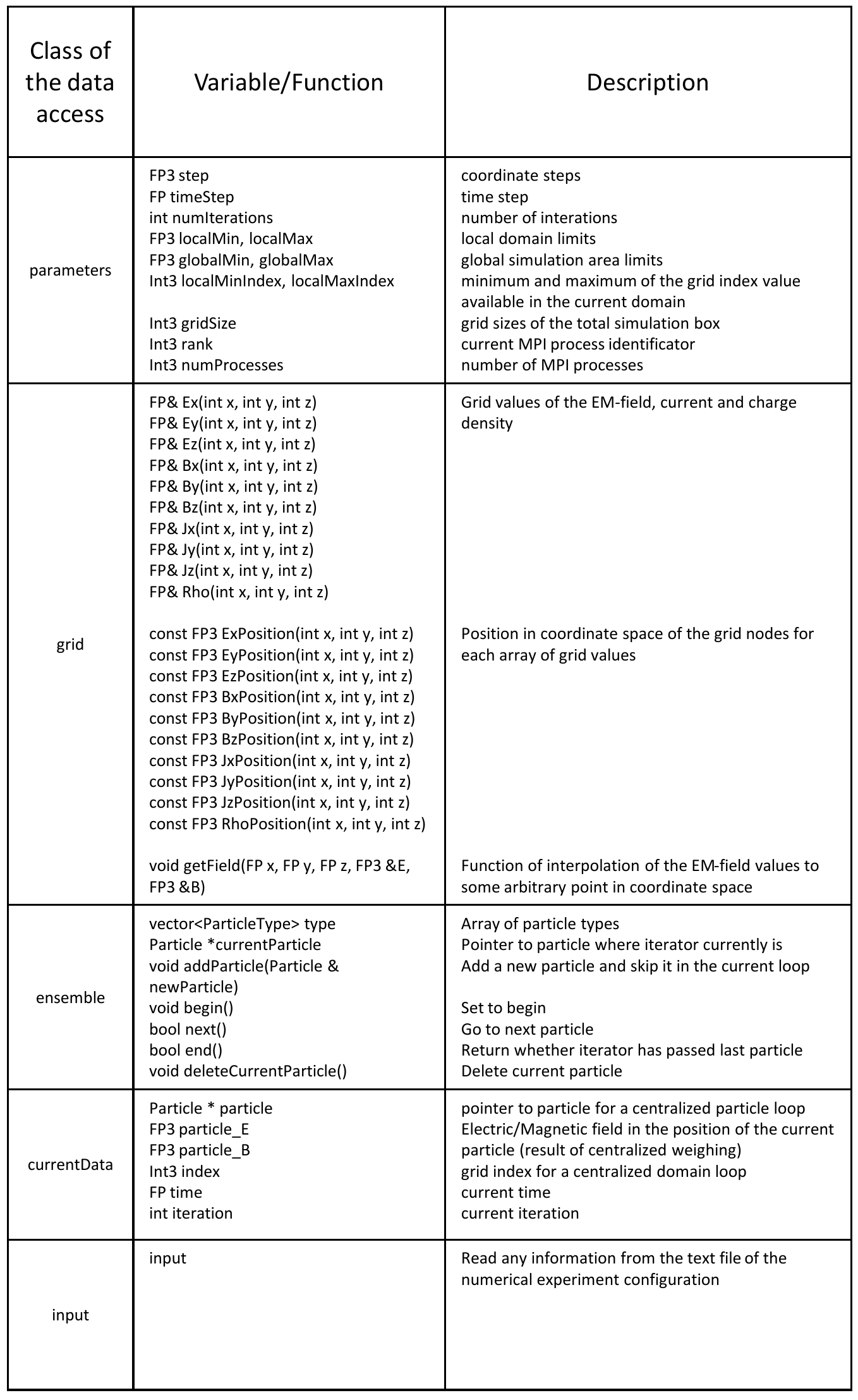}
\caption{Description of the PIC data access structure of the PIC-MDK interface. \texttt{FP} is a type for floating point operations (can be either \texttt{float} or \texttt{double}). \texttt{FP3} and \texttt{Int3} are types for arrays of three variables.}
\label{picmdk_acc}
\end{figure}

The access to the PIC data is provided via functions of a special class, `\texttt{Controller}'. The pointer to the object (the only for each domain) of this class is included in the class {Module} and is accessible by each module. The PIC data is structured in the form of five sub-objects: \texttt{parameters}, \texttt{grid}, \texttt{ensemble}, \texttt{currentData} and \texttt{input}. The listed objects are implemented in the form of separate classes described in Fig.~\ref{picmdk_acc}. \\

As one can see from Fig.~\ref{picmdk_acc}, access to the grid values is provided in a very general way, so that an implementation of the module is independent on whether the grid nodes of the electric and magnetic fields are defined in the shifted positions (Yee mesh \cite{yee.1966}) or in the same positions (spectral method). Also, to hide a half-step temporal shift for various variables (as is typical in leap-frog approaches), we symmetrize the PIC code relative to the time step. For the particles motion this is provided by the Boris method. Instead of a full step due to the electric field and, then, a full step due to the magnetic field, it assumes half-step due to the electric field, then the full step due to the magnetic field, and finally again the half-step due to the electric field. A similar symmetrization can be applied to the FDTD method and its modifications (see PICADOR code description, Sec.~\ref{SECTION:PICADOR}) and to the spectral method (see ELMIS code description, Sec.~\ref{SECTION:ELMIS}). The specificity of the data allocation for the particles is also abstracted with the \texttt{Ensemble} class providing internal iterator for the particles of a domain. \\

The modules are supposed to act at one or multiple pre-specified points within the main computational loop. These points are called \texttt{ports} and are placed at all necessary places of the computational loop, such as after evolving the EM-field or before handling a super-particle within the loop over them. The module developer should define which of the ports are required (through overloading the specialized function \texttt{ vector<PortID> getPorts()}) and to write a code for each of the required ports (through overloading the specialized function \texttt{void run(PortID portID)}). \\

For developers of a PIC implementation, adapting the code for PIC-MDK interface mainly implies 1) implementing the classes of PIC data access according to the description in Fig.~\ref{picmdk_acc}, 2) integrating the ports into the code and 3) adding the classes \texttt{Controller} and \texttt{Module} into the code structure. To the best of our knowledge, apart from a number of minor modifications, the PIC-MDK interface is sufficient for implementing all the described modifications. The PIC codes ELMIS~\cite{gonoskov.icnsp.2009, korzhimanov.icnsp.2011} and PICADOR~\cite{bastrakov.jcs.2012} have been adapted to the PIC-MDK interface and are being used successfully~\cite{gonoskov.prl.2013, gonoskov.prl.2014}. These two codes are reviewed briefly below.\\

\subsection{ELMIS: a parallel PIC code based on dispersion-free spectral-rotational Maxwell solver}\label{SECTION:ELMIS}

The code ELMIS (Extreme Laser-Matter Interaction Simulator) is a fully parallelized, 3D/2D PIC implementation based on the spectral method for solving Maxwell's equations and an original parallel algorithm for fast Fourier transform~\cite{gonoskov.nn.2004}. First presented at~\cite{gonoskov.icnsp.2009} in 2009, ELMIS code \cite{gonoskov.phd.2013} has been successfully applied to a number of studies in the field of laser-plasma interactions, including ion acceleration~\cite{gonoskov.prl.2009}, laser wakefield acceleration~\cite{soloviev.nima.2011, burza.prstab.2013} and higher harmonic generation at solids~\cite{gonoskov.pre.2011}. The code is based on an original \textit{spectral-rotational Maxwell solver} (for more details see \cite{gonoskov.phd.2013} Sect.~6.2.); this is a spectral method~\cite{dawson.rmp.1983} modification, which has no dispersion errors due to either grid step (see \cite{yu.arxiv.2013}) or time step (see \cite{vay.jcp.2013}). Just as for the pseudo-spectral algorithm described in~\cite{vay.jcp.2013}, our solver is based on an analytical solution for the evolution of EM-field Fourier harmonics, which provides the exact EM-field evolution for an arbitrary large time step. Furthermore, the use of a Fourier transform allows for charge conservation, via direct solution of Poisson's equation at each time step, and various possibilities for spectral filtering that can reduce numerical (statistical) noise and suppress numerical instabilities.  

\subsection{PICADOR code}\label{SECTION:PICADOR}

PICADOR \cite{bastrakov.jcs.2012, bastrakov.lncs.2014} is a fully parallel 3D PIC implementation capable of running on heterogeneous cluster systems with CPUs, GPUs, and Xeon Phi coprocessors. The features of PICADOR include FDTD and NDF field solvers, Boris particle pusher, CIC and TSC particle form factors, Esirkepov current deposition, ionization, moving frame, and dynamic load balancing. Each MPI process handles a part of simulation area (domain) using a multicore CPU via OpenMP, a GPU via CUDA, or a Xeon Phi coprocessor. All MPI exchanges occur only between processes handling neighboring domains. \\

A key aspect of high-performance implementation of the Particle-in-Cell method is to obtain an efficient memory access pattern during the most intense Particle--Grid operations: field interpolation and current deposition. We store particles of each cell in a separate array and process particles in a cell-by-cell order. This scheme helps to improve memory locality and allows vectorization of the particle loops. The particle storage scheme is encapsulated using the Ensemble class from the MDK, so that modules operating in the MDK-level abstractions for particle traversal automatically benefit from memory-friendly particle layout. However, the OpenMP-parallelized parts of PICADOR have to be coded explicitly using lower-level data structures. \\

PICADOR demonstrates performance and scaling on shared memory comparable to state-of-the-art implementations \cite{fonseca.ppcf.2013,decyk.cpc.2014}. On a simulation of a dense plasma with CIC field interpolation and current deposition in double precision, PICADOR achieves 12 nanoseconds per particle update on an 8-core Intel Xeon E5-2690 CPU with a 99\% strong scaling efficiency on shared memory. \\

The implementation for Xeon Phi is essentially the same C++/OpenMP code as for CPUs with a minor difference in the compiler directives that control vectorization. A Xeon Phi 7110X coprocessor in native mode scores 8 nanoseconds per particle update on the same benchmark, thus outperforming the Xeon E5-2690 CPU by factor of 1.5. A heterogeneous Xeon + Xeon Phi combination, with one process running on the processor and another one on the coprocessor, achieves 6 nanoseconds per particle update. However, other heterogeneous configurations, such as 2x Xeon + Xeon Phi or Xeon + 2x Xeon Phi, do not yield any performance benefit due to high MPI exchanges overhead. \\

The GPU implementation employs a variation on the widely used supercell technique \cite{burau.tps.2010} with a CUDA block processing the particles of a supercell. The main performance challenge in a GPU implementation is the current deposition, which requires reduction of the results of all threads in a block. We have two implementations of this operation: reduction in shared memory and reduction via atomic operations. The first one appears to be better on Fermi-generation GPUs, while the second is preferable on Kepler-generation GPUs, achieving 4x and 10x speedup over 8 CPU cores in single precision, respectively.

\bibliography{literature}
 \bibliographystyle{unsrt}

\end{document}